\documentclass[aps,prd,preprintnumbers,showpacs,showkeys,nofootinbib,
superscriptaddress,fleqn,floatfix,tightenlines]{revtex4-1}
\usepackage{amsmath,amsfonts,amssymb,amscd,amsxtra,amsthm}
\usepackage{graphicx}  % Include figure files
\usepackage{epstopdf}
\usepackage{dcolumn}  % Align table columns on decimal point
\usepackage{bm}          % bold math
\usepackage{slashed}
\usepackage[utf8]{inputenc} 
\usepackage{booktabs}
\usepackage[normalem]{ulem} % \sout{old text} for strikeout
\usepackage[dvipsnames]{xcolor} % For blue in-text comments and
\usepackage{tabularx}
\usepackage{enumitem}  
\usepackage{array} 
\usepackage{slashed}
\usepackage{tikz}
\usepackage{cleveref} 
\usepackage{multirow}
\renewcommand\sout{\bgroup \color{red} \ULdepth=-.5ex \ULset}

\makeatletter
\begin{document}  
\preprint{INHA-NTG-05/2018}
\title{Heavy baryons in a pion mean-field approach: A brief review}    
%--------------------------------------------------
\author{Hyun-Chul Kim}
\email[E-mail: ]{hchkim@inha.ac.kr}
\affiliation{Department of Physics, Inha University, Incheon 22212,
Republic of Korea}
\affiliation{School of Physics, Korea Institute for Advanced Study 
  (KIAS), Seoul 02455, Republic of Korea}

\date{\today}
\begin{abstract}
We review in this paper a series of recent works on properties of 
singly heavy baryons, based on a pion mean-field approach. 
In the limit of an infinitely heavy-quark mass, the heavy quark inside
a heavy baryon can be regarded as a static color source. In this
limit, a heavy baryon can be viewed as $N_c-1$ valence quarks bound by
the pion mean fields which are created self-consistently by the
presence of the $N_c$ valence quarks. We show that this mean-field
approach can successfully describe the masses and the magnetic moments
of the lowest-lying singly heavy baryons, using all the parameters
fixed in the light-baryon sector except for the hyperfine spin-spin
interactions. We also review a recent work on identifying the newly
found excited $\Omega_c$ baryons reported by the LHCb
Collaboration. We discuss possible scenarios to identify them. 
Finally, we give a future perspective on this pion mean-field approach. 
\end{abstract}
\maketitle
%----------------------------------
\section{Introduction}
%----------------------------------
An ordinary heavy baryon constitutes a pair of light quarks and a
heavy quark. Since the charm and bottom quarks are very heavy in
comparison with the light quarks, it is plausible to take the limit of
the infinitely heavy mass of the heavy quark, i.e. $m_Q\to \infty$. In
this limit, the physics of heavy baryons become simple. The spin of
the heavy quark is conserved, because of its infinitely heavy mass. It
results in the conservation of the total spin of light quarks:
$\bm{J}_L \equiv \bm{J}-\bm{J}_Q$, where $\bm{J}_L$, $\bm{J}_Q$, and
$\bm{J}$ denote the spin of the light-quark pair, that of the
heavy quark, and the total spin of the heavy baryon. This is called 
the heavy-quark spin symmetry that allows $\bm{J}_L$ to be a
good quantum number. Moreover, the physics is kept intact under the
placement of heavy quark flavors. This is called the heavy-quark
flavor symmetry~\cite{Isgur:1989vq, Isgur:1991wq,
  Georgi:1990um, Manohar:2000dt}. Then a heavy quark becomes static,
so that it can be considered as a static color source. Its importance
is only found in making the heavy baryon a color singlet, and
in giving higher-order contributions arising from $1/m_Q$ 
corrections. Consequently, the dynamics inside a heavy baryon is
mainly governed by the light quarks. 

\begin{figure}[htp]
\centering
\includegraphics[scale=0.35]{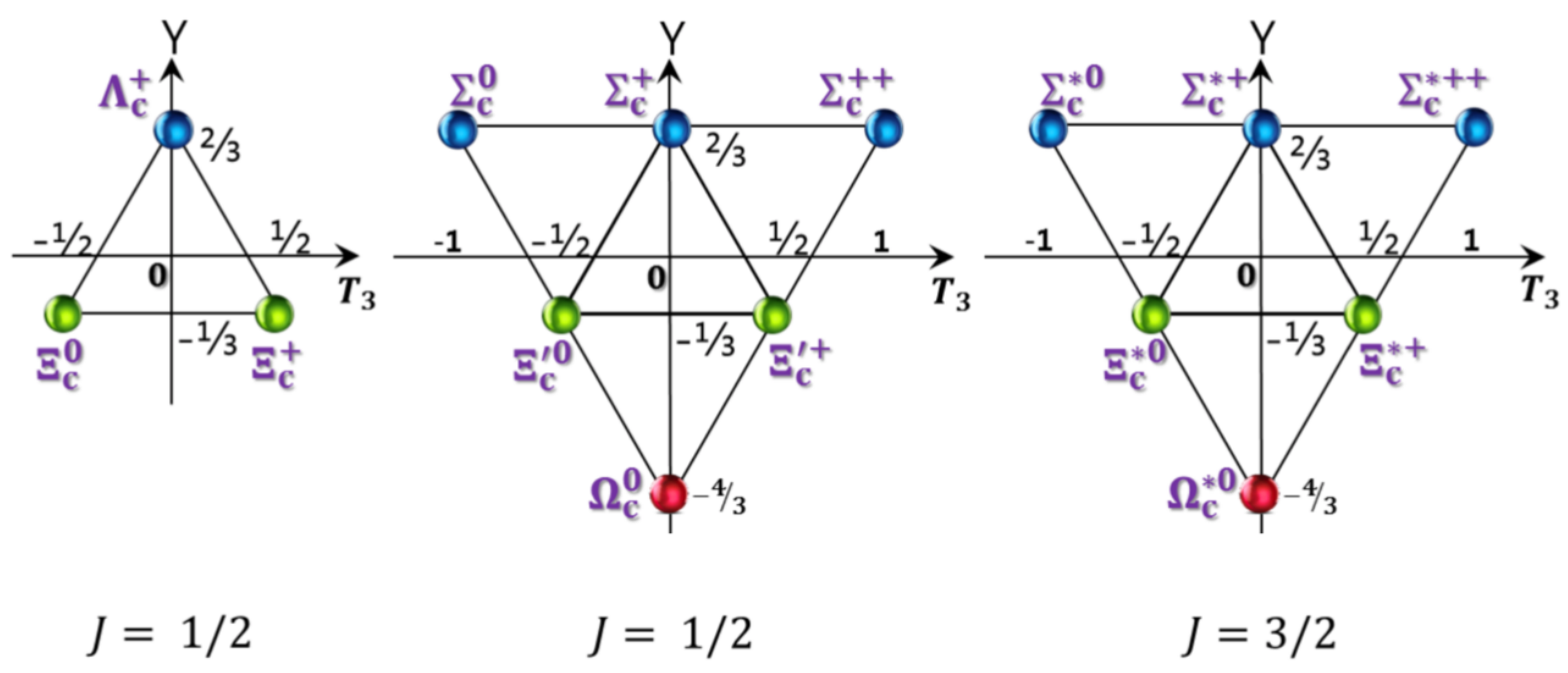}\qquad
\caption{The anti-triplet ($\overline{\bm{3}}$) and sextet ($\bm{6}$)
  representations of the lowest-lying heavy baryons. The left panel
  draws the weight diagram for the anti-triplet with the total spin
  $\frac{1}{2}$. The centered panel corresponds to that for the sextet
  with the total spin $1/2$ and the right panel depicts that
  for the sextet with the total spin $3/2$.} 
\label{fig:1}
\end{figure}

The flavor structure of the heavy baryon is also determined by
them. Since there are two light quarks inside the heavy baryon, we
have two different flavor  $SU_{\mathrm{f}}(3)$ irreducible
representations, i.e. $\bm{3}\otimes \bm{3}=\overline{\bm{3}} \oplus
\bm{6}$.  In the language of a quark model, the spatial part of the  
heavy-baryon ground state is symmetric due to the zero orbital angular
momentum, and the color part is totally antisymmetric. Since the
flavor anti-triplet ($\overline{\bm{3}}$) is antisymmrtric, the spin
state corresponding to {$\overline{\bm{3}}$} should be antisymmetric. Thus, the
baryons belonging to the anti-triplet should be $J_L=0$. Similarly, the
flavor-symmetric sextet ($\bm{6}$) should be symmetric in spin space,
i.e. $J_L=1$. This leads to the fact that the baryon antitriplet has
spin $J=1/2$, while the baryon sextet carries spin $J=1/2$ or $J=3/2$,
with the spin of the light-quark pair being coupled with the heavy
quark spin $J_Q=1/2$.  So, we can classify 15 different lowest-lying
heavy baryons as shown in Fig.~\ref{fig:1} in the case of charmed
baryons.   

Recently, there has been a series of new experimental data on the
spectra of heavy baryons~\cite{Aaltonen:2007ar, Chatrchyan:2012ni,
  Abazov:2008qm, Kuhr:2011up, Aaij:2012da, Aaij:2013qja, Aaij:2014esa,
  Aaij:2014lxa, Aaij:2014yka}, which renewed interest
in the physics of the heavy baryons. The lowest-lying singly heavy baryons
are now almost classified except for $\Omega_b ^\ast$. In the meanwhile, the
LHCb Collaboration has announced the first finding of two heavy
pentaquarks, $P_c(4380)$ and $P_c(4450)$ ~\cite{Aaij:2015tga,
  Aaij:2016phn, Aaij:2016ymb, Aaij:2016iza}. Very recently, the
five excited $\Omega_c$ baryons were reported~\cite{Aaij:2017nav}, among
which the four of them was confirmed by the Belle
experiment~\cite{Yelton:2017qxg}. Interestingly the two of the excited
$\Omega_c$s, i.e. $\Omega_c(3050)$ and $\Omega_c(3119)$, have very narrow
widths: $\Gamma_{\Omega_c(3050)}=(0.8\pm0.2\pm0.1)\,\mathrm{MeV}$ and
$\Gamma_{\Omega_c(3119)}=(1.1\pm0.8\pm0.4)\,\mathrm{MeV}$.   

While there is a great deal of theoretical approaches for the
description of heavy baryons, we will focus on a pion mean-field
approach in the present short review. This mean-field 
approach was first proposed by E. Witten in this seminal
papers~\cite{Witten:1979kh,Witten:1983}, where he asserted that in the
limit of the large number of colors ($N_c$) the nucleon can be
regarded as a bound state of $N_c$ \textit{valence} quarks in a pion
mean field with a hedgehog symmetry~\cite{Pauli:1942kwa,
  Skyrme:1961vq}. Since a baryon mass 
is proportional to $N_c$ whereas the quantum fluctuation around the
saddle point of the pion field is suppressed by $1/N_c$, the
mean-field approach is a rather plausible method for explaining
properties of baryons. The presence of $N_c$ \textit{valence} quarks
in this large $N_c$ limit, which consist of the lowest-lying baryons,    
produce the pion mean fields by which they are influenced
\emph{self-consistently}. This picture is very similar to a Hartree 
approximation in many-body theories. Witten also showed how to
construct the mean-field theory for the baryon schematically in
two-dimensional quantum chromodynamics (QCD). Though his idea was
criticized sometimes ago by S. Coleman~\cite{Coleman} because of its
technical difficulties, it is worthwhile to pursue it to see how far
we can describe the structure of the baryon in the pion mean-field
approach.   

The chiral quark-soliton model ($\chi$QSM)~\cite{Diakonov:1987ty,
  Christov:1995vm, Diakonov:1997sj} has been
constructed based on Witten's argument. The $\chi$QSM starts from the
effective chiral action (E$\chi$A) that was derived from the instanton
vacuum~\cite{Diakonov:1983hh, Diakonov:1985eg}. The E$\chi$A respects
chiral symmetry and its spontaneous breakdown, in which the essential
physics of the lowest-lying hadrons consists. One can derive the
classical energy of the nucleon by computing the nucleon correlation
function in Euclidean space, taking the Euclidean time to go to
infinity. Minimizing the classical energy self-consistently in the
large $N_c$ limit with the $1/N_c$ meson quantum fluctuations
suppressed, we obtain the classical mass and the self-consistent
profile function of the chiral soliton. While we ignore the $1/N_c$
quantum fluctuations around the saddle point of the soliton field, we
need to take into account the zero modes that do not change the
soliton energy. Since the soliton with hedgehog symmetry is not
invariant under translational, rotational and isotopic
transformations, we impose these symmetry properties on the
soliton and obtain a completely new solution with the same classical
energy. Because of the hedgehog symmetry, an  
$\mathrm{SU(2)}$ soliton needs to be embedded into the isospin
subgroup of the flavor
$\mathrm{SU(3)}_{\mathrm{f}}$~\cite{Witten:1983}, which was   
already utilized by various chiral soliton
models~\cite{Guadagnini:1983uv, Mazur:1984yf, Jain:1984gp}. This
collective quantization of the chiral soliton leads to the collective
Hamiltonian with effects of flavor $\mathrm{SU(3)}_{\mathrm{f}}$
symmetry breaking. The $\chi$QSM has one salient feature: the  
right hypercharge is constrained to be $Y'=N_c/3$ imposed by the $N_c$ 
valence quarks. This right hypercharge selects allowed representations
of light baryons such as the baryon octet ($\bm{8}$), the decuplet
($\bm{10}$), etc. The $\chi$QSM was successfully applied to the
properties of the lowest-lying light baryons such as the mass
splittings~\cite{Blotz:1992pw, Yang:2010fm}, the form
factors~\cite{Kim:1995mr, Silva:2001st, Ledwig:2010tu},
the magnetic moments~\cite{Kim:1995ha, Wakamatsu:1996xm, Kim:1997ip,
  Kim:2005gz}, hyperon 
semileptonic decays~\cite{Ledwig:2008ku, Yang:2015era}, 
parton distributions~\cite{Diakonov:1996sr, Wakamatsu:2003wg},
transversities of the nucleon~\cite{Kim:1995bq, Kim:1996vk,
  Schweitzer:2001sr}, generalized parton
distributions~\cite{Goeke:2001tz}, and so on.     

\begin{figure}[htp]
\centering
\includegraphics[scale=0.25]{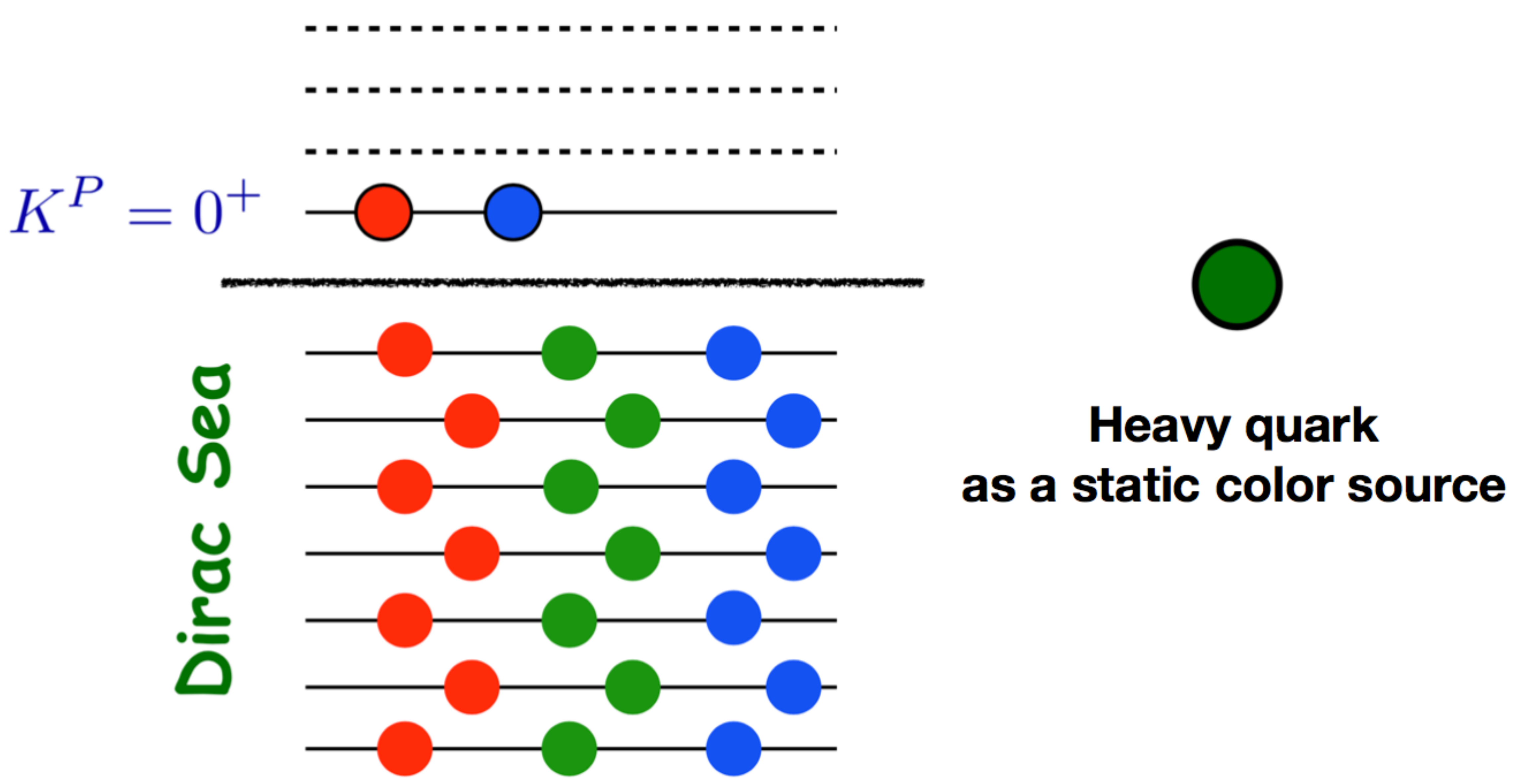}
\caption{Schematic picture of a heavy baryon. The $N_c-1$ valence
  quarks are filled in the lowest-lying valence level $K^P=0^+$ with
  the heavy quark stripped off. $K^P$ denotes the grand spin which we
  will explain later and  $P$ is the corresponding parity of the
  level. The presence of the valence quarks will interact with the sea
  quarks filled in the Dirac sea each other. This interaction will
  bring about the pion mean field.} 
\label{fig:2}
\end{figure}
Very recently, Ref.~\cite{Yang:2016qdz} extended a mean-field 
approach to describe the masses of singly heavy baryons, being  
motivated by Ref.~\cite{Diakonov:2010tf}. A singly heavy baryon
constitutes a heavy baryon and $N_c-1$ light valence quarks (see
Fig.~\ref{fig:2}). In the limit of $m_Q\to\infty$, the heavy quark can
be considered as a static color source. Thus, the dynamics inside a
heavy baryon is governed by the $N_c-1$ valence quarks. The presence
of the $N_c-1$ valence quarks will produce the pion mean fields as in
the case of the light baryons. However, there is one very significant
difference.   
the constraint right hyper charge is taken to be $Y' =(N_c-1)/3$ and
allows the lowest-lying representations: the baryon anti-triplet
($\overline{\bm{3}}$), the baryon sextet ($\bm{6}$), the baryon
anti-decapentaplet ($\overline{\bm{15}}$). The model reproduced
successfully the mass splitting of the baryon anti-triplet and sextet
in both the charm and bottom sectors. In addition, the mass of the
$\Omega_b^\ast $ baryon, which has not yet found, was predicted. The model
was further extended by including the second-order perturbative
corrections of flavor $SU_{\mathrm{f}}(3)$ symmetry
breaking~\cite{Kim:2018xlc}. The magnetic moments 
baryons~\cite{Yang:2018uoj} and electromagnetic form
factors~\cite{Kim:2018nqf} of the singly heavy baryons were also
studied within the same framework. The $\chi$QSM was also used to
interpret the five $\Omega_c$ baryons newly found by the LHCb
Collaboration~\cite{Kim:2017jpx, Kim:2017khv}. Within the present
framework, two of the $\Omega_c$s with the smaller widths are
classified as the members of the baryon $\overline{\bm{15}}$, whereas
all other $\Omega_c$'s belong to the excited baryon sextet. The widths
were quantitatively well reproduced without any free parameter. In the
present work, we will review briefly these recent investigations on the singly
heavy baryons. 

We sketch the present work as follows: In Section
II, we review the general formalism of the $\chi$QSM for singly heavy
baryons. In Section III, we examine the mass splittings of the heavy
baryons, emphasizing the discussion of the effects of
$\mathrm{SU(3)}_{\mathrm{f}}$ breaking. In Section IV, we discuss the
recent results of the magnetic moments and electromagnetic form
factors of the heavy baryons. In Section V, we briefly introduce a
theoretical interpretation of the excited $\Omega_c$ baryons found by the
LHCb, based on the present mean-field approach. The final Section is
devoted to the conclusions and outlook.

\section{The chiral quark-soliton model for singly heavy baryons} 
In the present approach, a heavy baryon is considered as a bound state
of the $N_c-1$ valence quarks in the pion mean field with a heavy
quark stripped off from the valence level. Thus, the correlation
function of the heavy baryon can be expressed in terms of the $N_c-1$
valence quarks
\begin{align}
\Pi_{B}(0, T) = \langle J_B (0, T/2) J_B^\dagger
  (0,-T/2) \rangle_0 = \frac{1}{Z}\int  \mathcal{D} U
  \mathcal{D}\psi^\dagger 
  \mathcal{D}\psi J_B(0,T/2) J_B^\dagger (0,-T/2)
  e^{\int d^4 x\,\psi^\dagger (i\rlap{/}{\partial} + i
  MU^{\gamma_5}+ i \hat{m})\psi}  ,   
\label{eq:corr1}
\end{align}
where $J_B$ denotes the light-quark current with the $N_c-1$ 
light quarks for a heavy baryon $B$
\begin{align}
J_B(\bm{x}, t) = \frac1{(N_c-1)!}
  \varepsilon^{\beta_1\cdots\beta_{N_c-1}} \Gamma_{J'J_3',TT_3}^{\{f\}}
  \Psi_{\beta_1f_1}(\bm{x}, t) \cdots \Psi_{\beta_{N_c-1}f_{N_c-1}}
  (\bm{x}, t).      
\end{align}
$\beta_i$ stand for color indices and
$\Gamma_{J'J_3',TT_3}^{\{f_1\cdots f_{N_c-1}\}}$ represents a matrix 
with both flavor and spin indices. $J'$ and $T$ are the spin and
isospin of the heavy baryon, respectively. $J_3'$ and $T_3$ are their
third components, respectively. The notation $\langle \cdots \rangle_0$ in
Eq.~(\ref{eq:corr1}) is the vacuum expectation value, $M$
the dynamical quark mass, and the chiral field $U^{\gamma_5}$ is
defined as    
\begin{align}
U^{\gamma_5} = U\frac{1+\gamma_5}{2} + U^\dagger \frac{1-\gamma_5}{2}   
\end{align}
with 
\begin{align}
U = \exp(i\pi^a \lambda^a).  
\end{align}
Here, $\pi^a$ represents the pseudo-Goldstone boson field and
$\hat{m}$ denotes the flavor matrix of the current quarks, written as
$\hat{m}=\mathrm{diag}(m_{\mathrm{u}},\,m_{\mathrm{d}},\,m_{\mathrm{s}})$. We 
assume isospin symmetry, i.e. $m_{\mathrm{u}}=m_{\mathrm{d}}$. Since
the strange current quark mass is small enough, we will treat it
perturbatively.   

Integrating over the quark fields, we derive the correlation function
as 
\begin{align}
\Pi_{B}(0, T) =
  \frac{1}{Z}\Gamma_{J'J_3',TT_3}^{\{f\}}\Gamma_{J'J_3',TT_3}^{\{g\}*} \int
  \mathcal{D} U \prod_{i=1}^{N_c-1}  \left\langle 0,T/2\left|
  \frac1{D(U)} \right|0,-T/2\right\rangle
  e^{-S_{\mathrm{eff}}(U)}, 
\label{eq:corr2}
\end{align}  
where the single-particle Dirac operator $D(U)$ is defined as 
\begin{align}
D(U) = i\gamma_4 \partial_4 + i\gamma_k \partial_k + i MU^{\gamma_5} +
  i \hat{m}   
\end{align}
and $S_{\mathrm{eff}}$ is the effective chiral action written
as 
\begin{align}
S_{\mathrm{eff}} = -N_c \mathrm{Tr}\log D(U).  
\label{eq:effecXac}
\end{align}
Equation~(\ref{eq:corr2}) can be schematically depicted as
Fig.~\ref{fig:2}. It consists of two different terms: The first and
second ones are respectively called the \textit{valence-quark
  contribution} and \textit{sea-quark contribution} within the
$\chi$QSM.  
\begin{figure}[htp]
\centering
\includegraphics[scale=0.3]{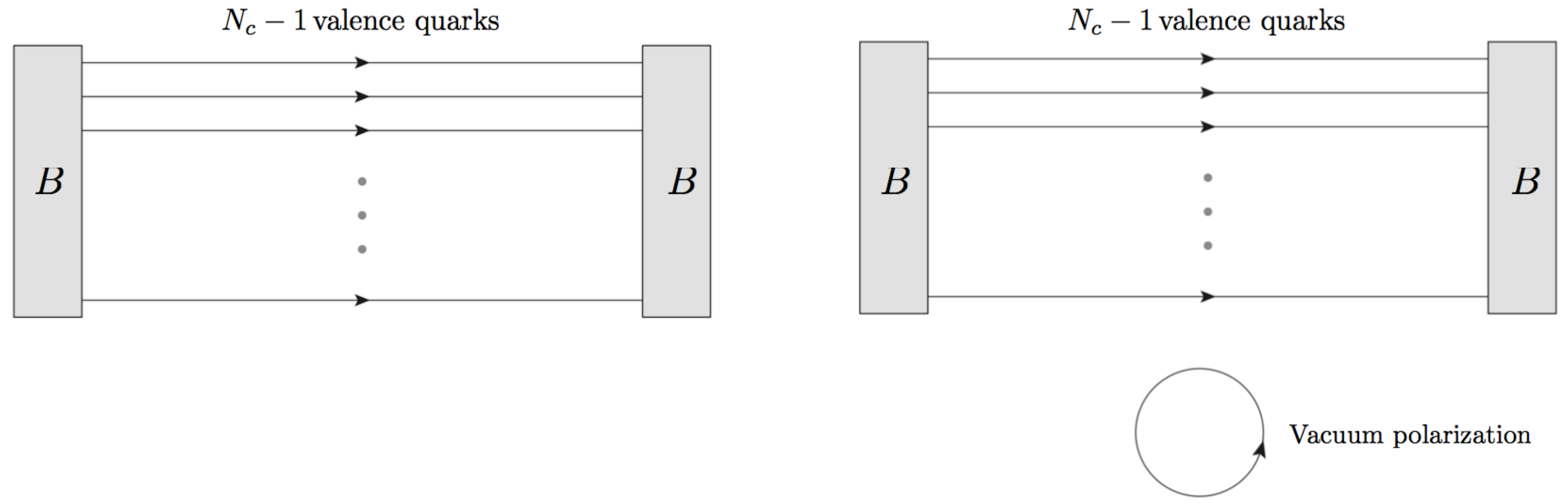}\qquad
\caption{Correlation function for a heavy baryon} 
\label{fig:3}
\end{figure}
When the Euclidean time $T$ is taken from $-\infty$ to $\infty$, 
the correlation function picks up the ground-state
energy~\cite{Diakonov:1987ty, Christov:1995vm} 
\begin{align}
\lim_{T\to\infty} \Pi_B(T) \sim \exp[-\left\{(N_c-1) E_{\mathrm{val}} +
  E_{\mathrm{sea}}\right\} T],   
\end{align}
where $E_{\mathrm{val}}$ and $E_{\mathrm{sea}}$ the valence and
sea quark energies. Minimizing self-consistently the energies around
the saddle point of the chiral field $U$
\begin{align}\label{eq:sol}
\left.\frac{\delta}{\delta U}[ (N_c-1) E_{\mathrm{val}} +
  E_{\mathrm{sea}}]\right|_{U_c} = 0,    
\end{align}
we get the classical soliton mass 
\begin{align}\label{eq:solnc}
M_{\mathrm{sol}} = (N_c-1) E_{\mathrm{val}}(U_c) + E_{\mathrm{sea}}(U_c).  
\end{align}
Note that a singly heavy baryon has a heavy quark, so its classical
is expressed as the sum of the classical and heavy-quark masses
\begin{align}
M_{\mathrm{cl}} = M_{\mathrm{sol}} + m_Q.  
\label{eq:classical_mass}
\end{align}
We want to mention that $m_Q$ is the \textit{effective} heavy quark
mass that is different from that of QCD and will be absorbed
in the center mass of each representation. 

The rotational excitations of the soliton with $N_c-1$ valence quarks
will produce the lowest-lying heavy baryons. To keep the hedgehog
symmetry, the SU(2) soliton $U_c(\bm{r})$ will be embedded into
SU(3)~\cite{Witten:1983}  
\begin{align}
U(\bm{r}) = \begin{pmatrix}
U_c (\bm{r}) & 0 \\
0 & 1
\end{pmatrix}.
\label{eq:embed}
\end{align}
As mentioned in Introduction, we consider explicitly the rotational zero
modes.  Assuming that the soliton $U(\bm{r})$ in Eq.(\ref{eq:embed})
rotates slowly, we apply the rotation matrix $A(t)$ in
$\mathrm{SU}_{\mathrm{f}}(3)$ space    
\begin{align}
U(\bm{r},\,t) = A(t) U(\bm{r}) A^\dagger (t).  
\end{align}
Then, we can derive the collective Hamiltonian for heavy
baryons
\begin{align}
H =& H_{\mathrm{sym}} + H^{(1)}_{\mathrm{sb}} +  H^{(2)}_{\mathrm{sb}},
\end{align}
where $H_{\mathrm{sym}}$ represents the flavor SU(3) symmetric part, 
$H^{(1)}_{\mathrm{sb}}$ and $H^{(2)}_{\mathrm{sb}}$ the
SU(3) symmetry-breaking parts respectively to the first and second
orders. $H_{\mathrm{sym}}$ is expressed
as  
\begin{align} 
H_{\mathrm{sym}}=M_{\mathrm{cl}}+\frac{1}{2I_{1}}\sum_{i=1}^{3}
\hat{J}^{2}_{i} +\frac{1}{2I_{2}}\sum_{a=4}^{7}\hat{J}^{2}_{a},  
\end{align} 
where $I_{1}$ and $I_{2}$ are the moments of inertia of the
soliton and the operators  $\hat{J}_{i}$ denote the SU(3) 
generators. We get the eigenvalue of the quadratic Casimir operator
$\sum_{i=1}^8 J_i^2$ in the $(p,\,q)$ representation, given as    
\begin{align}
C_2(p,\,q) = \frac13 \left[p^2 +q^2 + pq + 3(p+q)\right],   
\label{eq:Casimir}
\end{align}
which leads to the eigenvalues of $H_{\mathrm{sym}}$ 
\begin{align} 
E_{\mathrm{sym}}(p,q) = M_{\mathrm{cl}}+ \frac{1}{2I_{1}} J_L(J_L+1) 
+\frac{1}{2I_{2}}\left[C_2(p,\,q) - J_L(J_L+1)\right] 
-\frac{3}{8I_{2}} {Y'}^{2}.
\label{eq:RotEn}
\end{align} 
The right hypercharge $Y'$ is constrained by the $N_c-1$ valence
quarks inside a singly heavy baryon, i.e. $Y'=(N_c-1)/3$. The
corresponding collective wave functions of the singly heavy baryon is
then obtained as  
\begin{align} 
\psi_B^{({\mathcal{R}})}(JJ_3,J_L;A)=
\sum_{m_{3}=\pm1/2}C^{J J_3}_{J_{Q} m_{3} J_L 
J_{L3}} \chi_{m_{3}} \sqrt{\mathrm{dim}(p,\,q)}
(-1)^{-\frac{ Y'  }{2}+{J}_{L3}}
  D^{(\mathcal{R})\ast}_{(Y,T,T_3)(Y'  ,J_L,-J_{L3})}(A),  
\label{eq:waveftn}
\end{align} 
where 
\begin{align}
\mathrm{dim}(p,\,q) = (p+1)(q+1)\left(1+\frac{p+q}{2}\right).  
\end{align}
 $J$ and $J_3$ in Eq.~(\ref{eq:waveftn}) are the spin angular
 momentum and its third component of the heavy baryon, respectively.   
$J_L$ and $J_{Q}$ represent the soliton spin and
heavy-quark spin, respectively. ${J_{L3}}$ and ${m_{3}}$ are the
corresponding third components, respectively. Since the spin operator
for the heavy baryon is given as 
\begin{align}
\label{eq:quantum}
\bm{J}=\bm{J}_{Q}+\bm{J}_L,
\end{align} 
the relevant Clebsch-Gordan coefficients appear in
Eq.(\ref{eq:waveftn}). The SU(3) Wigner $D$ function in
Eq.(\ref{eq:waveftn}) means just the wave-function for the quantized
soliton with the $N_c-1$ valence quarks, and $\chi_{m_3}$
is the Pauli spinor for the heavy quark. $\mathcal{R}$ designates a
SU(3) irreducible representation corresponding to $(p,\,q)$. 
Since the soliton is coupled to the heavy quark, we finally obtain the
three lowest-lying representations illustrated in
Fig.~\ref{fig:1}. In the limit of $m_Q\to\infty$, the two sextet
representations are degenerate. One needs to introduce a 
hyperfine spin-spin interaction to lift this degeneracy. As will be
discussed soon, this hyperfine interaction will be determined by using
the experimental data on the masses of heavy baryons.    

\begin{figure}[htp]
\centering
\includegraphics[scale=0.2]{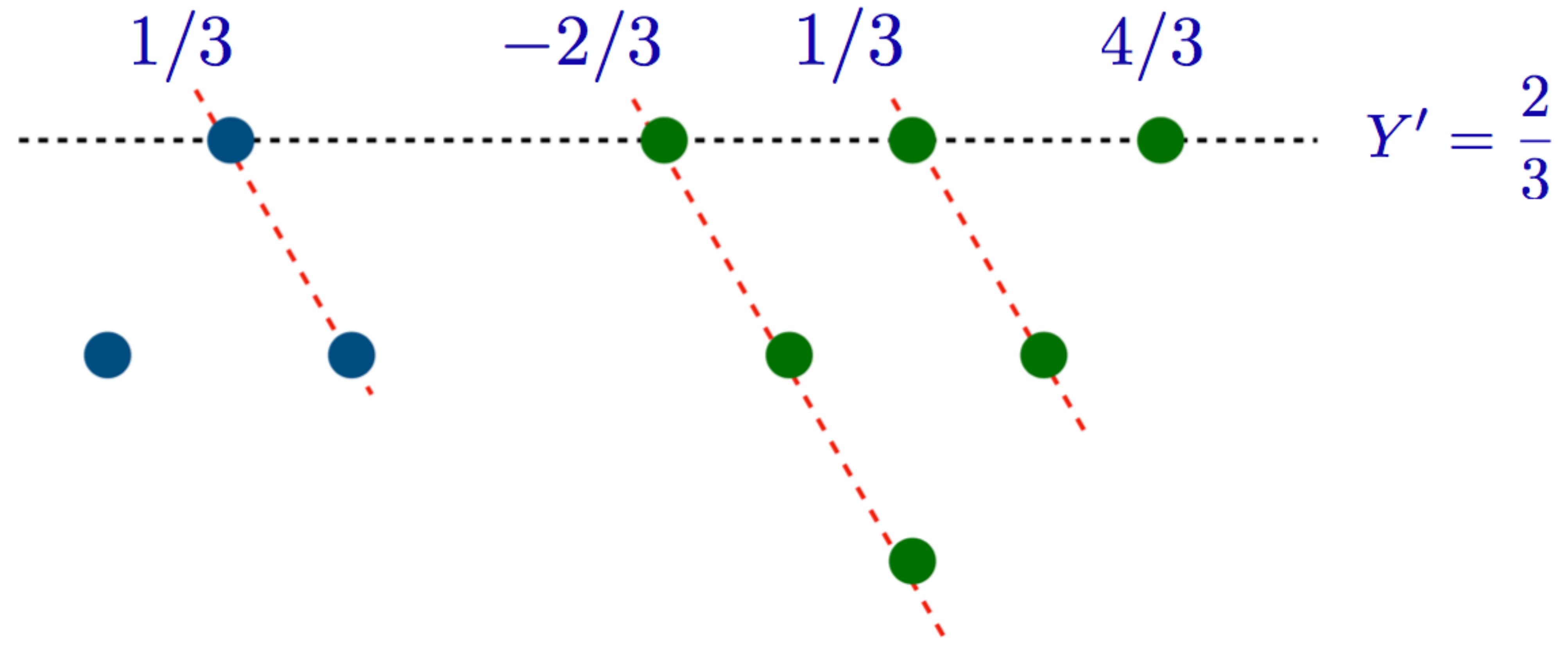}\qquad
\caption{The baryon anti-triplet has the  $J_{L}=T=0$ state with $Y'=2/3$ whereas
  the baryon sextet contains the  $J_{L}=T=1$ state with $Y'=2/3$.}
\label{fig:4}
\end{figure}

In the present zero-mode quantization scheme, we find the following
the two important selection rule. The allowed SU(3) representations
must contain states with $Y'=(N_c-1)/3$ and the isospin $\bm{T}$ of
the states with $Y'=(N_c-1)$/3 are coupled with the soliton so that we
have a singlet $\bm{K}=\bm{T}+\bm{J}_{L}=\bm{0}$, where $\bm{K}$ is called 
the grand spin. The lowest-lying heavy baryons have the grand spin
$K=0$, that is, we must have always $J_{L}=T$ with $Y'=(N_c-1)/3$ for the
ground-state heavy baryons as shown in fig.~\ref{fig:4}. 

An observable of the heavy baryon can be expressed in general as a
three-point correlation function
\begin{align}
\langle B,\,p'| J_\mu(0) |B,\,p\rangle  &= \frac1{\mathcal{Z}}
  \lim_{T\to\infty} \exp\left(i p_4\frac{T}{2} - i p_4'
  \frac{T}{2}\right) \int d^3x d^3y \exp(-i \bm{p}'\cdot \bm{y} + i
  \bm{p}\cdot \bm{x}) \cr
& \hspace{-1cm} \times \int \mathcal{D}U\int \mathcal{D} \psi \int
  \mathcal{D} \psi^\dagger J_{B}(\bm{y},\,T/2) \psi^\dagger(0)
  \gamma_4 \Gamma \mathcal{O} \psi(0) J_B^\dagger (\bm{x},\,-T/2)
  \exp\left[-\int d^4 z   \psi^\dagger iD(U) \psi\right],
  \label{eq:3corrftn}
\end{align}
where $\Gamma$ and $\mathcal{O}$ represent respectively generic Dirac
spin and flavor matrices. Computing Eq.~(\ref{eq:3corrftn}), one can  
study heavy baryonic observables such as form factors, magnetic moments,
axial-vector constants, etc. For the detailed formalism, we refer to
Refs.~\cite{Kim:1995mr, Christov:1995vm}. 

\section{Mass splittings  of the singly heavy baryons}
We first discuss the mass splittings of the singly heavy baryons. In
order to obtain the mass splittings, one should include the
symmetry-breaking part of the collective
Hamiltonian~\cite{Blotz:1992pw, Christov:1995vm} 
\begin{align} 
H^{(1)}_{\mathrm{sb}} 
&= \overline{\alpha} D^{(8)}_{88}+ \beta \hat{Y}
+ \frac{\gamma}{\sqrt{3}}\sum_{i=1}^{3}D^{(8)}_{8i}
\hat{J}_{i},
\label{sb}
\end{align}
where
\begin{align} 
\overline{\alpha} = \left (-\frac{\overline{\Sigma}_{\pi N}}{3m_0}+\frac{
  K_{2}}{I_{2}}Y'  
\right )m_{\mathrm{s}},
 \;\;\;  \beta=-\frac{ K_{2}}{I_{2}}m_{\mathrm{s}}, 
\;\;\;  \gamma = 2\left (
  \frac{K_{1}}{I_{1}}-\frac{K_{2}}{I_{2}}   \right ) m_{\mathrm{s}}.
\label{eq:alphaetc}
\end{align}
The parameters $\overline{\alpha}$, $\beta$, and
$\gamma$ are the essential ones in determining the masses of the
lowest-lying singly heavy baryons, which are expressed in terms of the
moments of inertia $I_{1,\,2}$ and $K_{1,\,2}$. However, we do not
need to fit them, since they are related to $\alpha$,
$\beta$, and $\gamma$ in the light-baryon sector. 
The valence parts are only different from those in the light baryon
sector by the color factor $N_c-1$. So, we need to replace $N_c$ by $N_c-1$
in the valence parts of all the relevant dynamical parameters
determined in the light-baryon sector. The valence part of
$\overline{\Sigma}_{\pi N}$ is just the $\pi N$ sigma term with
different $N_c$ factor: $\overline{\Sigma}_{\pi N} = (N_c-1)N_c^{-1}
\Sigma_{\pi    N}$, where $\Sigma_{\pi N} = (m_u+m_d)\langle N|
\bar{u} u +  \bar{d} d|N\rangle = (m_u+m_d) \sigma$. On the other
hand, the sea parts should be kept intact as in the light baryon
sector. 

The dynamical parameters $\alpha$, $\beta$ and $\gamma$ have been
fixed by using the experimental data on the baryon octet masses and
a part of the baryon decuplet and anti-decuplet masses with  
isospin symmetry breaking effects~\cite{Yang:2010id}. The
values of $\alpha$, $\beta$, and $\gamma$ have been obtained by
the $\chi^2$ fit~\cite{Yang:2010fm}
\begin{align}
\alpha  =  -255.03\pm5.82 \;{\rm MeV},\;\;\;
\beta  =  -140.04\pm3.20  \;{\rm MeV},\;\;\;
\gamma  =  -101.08\pm2.33 \;{\rm MeV},
\label{eq:abrNumber}
\end{align}
While $\beta$ and $\gamma$ are not required to be changed in the
heavy-baryon sector, $\alpha$ should be modified by
\begin{align}
  \label{eq:alpha}
\overline{\alpha} = \rho \alpha,
\end{align}
where $\rho=(N_c-1)/N_c$.
However, there is a caveat when one uses the values of
Eq.~\eqref{eq:abrNumber}. As mentioned above, only the valence parts
should be modified, while the scaling in Eq.~\eqref{eq:alpha} changes
the sea part too.  To compensate this we choose $\rho \approx 0.9$. If
one computes the parameters $\overline{\alpha}$, $\beta$, and $\gamma$
in a self-consistent way, we do not have this problem~\cite{Kim:2018xlc}.

Considering the first-order perturbative corrections of $m_{\mathrm{s}}$, 
one can express the masses of the singly heavy baryons in
representation $\mathcal{R}$ as  
\begin{align}
M_{B,\mathcal{R}}^Q = M_{\mathcal{R}}^Q + M_{B,\mathcal{R}}^{(1)}  
\label{eq:FirstOrderMass}
\end{align}
with
\begin{align}
M_{\mathcal{R}}^Q = m_Q + E_{\mathrm{sym}}(p,q).  
\end{align}
Here, $M_{\mathcal{R}}^Q$ is the center mass of a heavy baryon in
representation $\mathcal{R}$. $E_{\mathrm{sym}}(p,q)$ is the
eigenvalue energy of the symmetric part of the collective Hamiltonian
defined in Eq.~\eqref{eq:RotEn}. Note that the lower index $B$
designates a certain baryon in a specific representation
$\mathcal{R}$. The upper index $Q$ denotes either the charm sector
($Q=c$) or the bottom sector ($Q=b$). 
Then the center masses for the anti-triplet and sextet representations
are obtained as 
\begin{align}
M_{\overline{\bm{3}}}^Q = M_{\mathrm{cl}} + \frac1{2I_2}, \;\;\; 
M_{\bm{6}}^Q = M_{\overline{\bm{3}}}^Q +  \frac1{I_1},
\end{align}
where $M_{\mathrm{cl}}$ was defined in Eq.~\eqref{eq:classical_mass}.
The second term in Eq.~(\ref{eq:FirstOrderMass}), which arises from
the linear-order $m_{\mathrm{s}}$ corrections, is proportional to the
hypercharge of the soliton with the light-quark pair 
\begin{align}
M^{(1)}_{B,{\cal{R}}} = \langle B, {\cal{R}} | H_{\mathrm{sb}}^{(1)} 
| B, {\cal{R}} \rangle  = Y\delta_{{\cal{R}}},
\end{align}
where
 \begin{align} 
\delta_{\overline{\bm{3}}}=\frac{3}{8}\overline{\alpha}+\beta, \;\;\;\;
\delta_{\bm{6}}=\frac{3}{20}
  \overline{\alpha}+\beta-\frac{3}{10}\gamma.  
\label{eq:deltas}
\end{align}
Finally, we arrive at the expressions for the masses of the
lowest-lying baryon anti-triplet and sextet as follows 
\begin{align}
M_{B,\overline{\bm{3}}}^Q = M_{\overline{\bm{3}}}^Q  +
                     Y \delta_{\overline{\bm{3}}} ,\;\;\;
M_{B,\bm{6}}^Q =M_{\bm{6}}^Q  + Y  \delta_{\bm{6}}, 
  \label{eq:firstms}
\end{align}
with the linear-order $m_{\mathrm{s}}$ corrections taken into account. 

Since the baryon sextet with spin 1/2 and 3/2 are degenerate, we need
to remove the degeneracy by introducing the hyperfine spin-spin
interaction Hamiltonian~\cite{Zeldovich}. Typically, the hyperfine
Hamiltonian is written as 
\begin{align}
H_{LQ} = \frac{2}{3}\frac{\kappa}{m_{Q}\,M_{\mathrm{sol}}}\bm{J}_L
\cdot  \bm{J}_{Q}  
= \frac{2}{3}\frac{\varkappa}{m_{Q}}
\bm{J}_{L} \cdot \bm{J}_{Q}, 
\label{eq:ssinter}
\end{align}
where $\kappa$ stands for the flavor-independent hyperfine coupling.
$M_{\mathrm{sol}}$ has been incorporated into an unknown 
coefficient $\varkappa$ that will be fixed by using the experimental
data. . The Hamiltonian $H_{LQ}$ does not affect the 
$\overline{\bm{3}}$ states with $J_L=0$. On the other hand, 
the baryon sextet acquire additional contribution from $H_{LQ}$ which
bring about the splitting between different spin states 
\begin{align}
M_{B,{\bm{6}}_{1/2}}^{Q} =  
M_{B,\bm{6}}^Q\;-\;\frac{2}{3}\frac{\varkappa}{m_{Q}}, 
\;\;\;
M_{B,{\bm{6}}_{3/2}}^{Q}  = 
M_{B,\bm{6}}^Q\;+\;\frac{1}{3} \frac{\varkappa}{m_{Q}}, 
\label{eq:Csextet}
\end{align}
which leads to the splitting 
\begin{align}
M_{B,{\bm{6}}_{3/2}}^{Q}\;-\; M_{B,{\bm{6}}_{1/2}}^{Q}  =
  \frac{\varkappa}{m_{Q}}  . 
\label{eq:DCsextet}
\end{align}
The numerical values of $\varkappa/m_Q$ were determined by using the center
values of the masses of the baryon sextet~\cite{Yang:2016qdz} 
\begin{align}
\frac{\varkappa}{m_c} = (68.1\pm 1.1)\,\mathrm{MeV},\;\;\;
\frac{\varkappa}{m_b} = (20.3\pm 1.0)\,\mathrm{MeV}.
  \label{eq:kappavalue}
\end{align}
Note that $\varkappa$ is flavor-independent. So, knowing the ratio
$m_c/m_b$, one can extract the value of $\varkappa$ from
Eq.~\eqref{eq:kappavalue}.

We now present the numerical results of the masses of the heavy
baryons~\cite{Yang:2016qdz}. Using the values of $\overline{\alpha}$,
$\beta$, and $\gamma$, we can immediately determine the values of
$\delta_{\overline{\bm{3}}}$ and $\delta_{\bm{6}}$ defined in
Eq.~\eqref{eq:deltas} 
\begin{align}
\delta_{\overline{\bm{3}}} = (-203.8\pm 3.5)\,\mathrm{MeV},\;\;\;
\delta_{\bm{6}} = (-135.2\pm 3.3)\,\mathrm{MeV}.
\end{align}
Including the results of $\varkappa/m_c$ and $\varkappa/m_b$, we can
obtain the numerical results of the heavy baryon masses. 
In Table~\ref{tab:1} and Table~\ref{tab:2} the numerical results of
the charmed and bottom baryon masses are presented, respectively. They
are in good agreement with the experimental data taken from 
Ref.~\cite{PDG2017}. The mass of $\Omega_b^*$ is still experimentally
unknown. Thus, the prediction of its mass is given as 
\begin{align}
M_{\Omega_b^*} = (6095.0\pm 4.4)\,\mathrm{MeV}.  
\end{align}
The uncertainties in Tables~\ref{tab:1} and \ref{tab:2} are due to
those in $\overline{\alpha}$, $\beta$, $\gamma$, and $\varkappa/m_Q$. 
\begin{table}[htp]
\begin{centering}
\begin{tabular}{c|ccc}
\hline \hline
$\mathbf{\mathcal{R}}_{J}^{Q}$ 
& $B_{c}$ 
& Mass 
& Experiment
\tabularnewline[0.1em]
\hline 
\multirow{2}{*}{$\mathbf{\overline{3}}_{1/2}^{c}$} 
& $\Lambda_{c}$ 
& $2272.5 \pm 2.3$
& $2286.5 \pm 0.1$
\tabularnewline
& $\Xi_{c}$ 
& $2476.3 \pm 1.2$
& $2469.4 \pm 0.3$
\tabularnewline
\hline 
\multirow{3}{*}{$\mathbf{6}_{1/2}^{c}$} 
& $\Sigma_{c}$ 
& $2445.3 \pm 2.5$
& $2453.5 \pm 0.1$
\tabularnewline
& $\Xi_{c}^{\prime}$ 
& $2580.5 \pm 1.6$
& $2576.8 \pm 2.1$
\tabularnewline
& $\Omega_{c}$ 
& $2715.7 \pm 4.5$
& $2695.2 \pm 1.7$
\tabularnewline
\hline 
\multirow{3}{*}{$\mathbf{6}_{3/2}^{c}$} 
& $\Sigma_{c}^{\ast}$ 
& $2513.4 \pm 2.3$
& $2518.1 \pm 0.8$
\tabularnewline
& $\Xi_{c}^{\ast}$ 
& $2648.6 \pm 1.3$
& $2645.9 \pm 0.4$
\tabularnewline
& $\Omega_{c}^{\ast}$ 
& $2783.8 \pm 4.5$
& $2765.9 \pm 2.0$
\tabularnewline
\hline \hline
\end{tabular}
\par\end{centering}
\caption{The numerical results of the charmed baryon masses in 
  comparison with the experimental data~\cite{PDG2017}.}
\label{tab:1}
\end{table}

\begin{table}[htp]
\centering{}\vspace{3em}%
\begin{tabular}{c|ccc}
\hline\hline 
$\mathbf{\mathcal{R}}_{J}^{Q}$ 
& $B_{b}$ 
& Mass
& Experiment
\\
\hline 
\multirow{2}{*}{$\mathbf{\overline{3}}_{1/2}^{b}$} 
& \textcolor{black}{$\Lambda_{b}$} 
& $5599.3 \pm 2.4 $
& $5619.5 \pm 0.2$ 
 \tabularnewline
& \textcolor{black}{$\Xi_{b}$} 
& $5803.1 \pm 1.2 $
& $5793.1 \pm 0.7 $  
 \tabularnewline
\hline 
\multirow{3}{*}{$\mathbf{6}_{1/2}^{b}$} 
& \textcolor{black}{$\Sigma_{b}$} 
& $5804.3 \pm 2.4 $
& $5813.4 \pm 1.3$  
\tabularnewline
& \textcolor{black}{$\Xi_{b}^{\prime}$} 
& $5939.5 \pm 1.5 $
& $5935.0 \pm 0.05$ 
\tabularnewline
& \textcolor{black}{$\Omega_{b}$} 
& $6074.7 \pm 4.5 $
& $6048.0 \pm 1.9$  
 \tabularnewline
\hline 
\multirow{3}{*}{$\mathbf{6}_{3/2}^{b}$} 
& \textcolor{black}{$\Sigma_{b}^{\ast}$} 
& $5824.6 \pm 2.3 $
& $5833.6 \pm 1.3$  
\tabularnewline
&\textcolor{black}{$\Xi_{b}^{\ast}$} 
& $5959.8 \pm 1.2 $
& $5955.3 \pm 0.1$ 
 \tabularnewline
& \textcolor{black}{$\Omega_{b}^{\ast}$} 
& $6095.0 \pm 4.4 $
& $-$
\tabularnewline
\hline \hline
\end{tabular}
\caption{The results of the masses of the bottom baryons in comparison
with the experimental data~\cite{PDG2017}.}
\label{tab:2}
\end{table}

\section{Magnetic moments of heavy baryons} 
In this Section, we briefly summarize a recent work on the magnetic
moments of the heavy baryons~\cite{Yang:2018uoj}.
Starting from Eq.~\eqref{eq:3corrftn}, one can derive the general
expressions of the collective operator for the magnetic moments   
\begin{align}
  \label{eq:MagMomOp}
 \hat{\mu} = \hat{\mu}^{(0)}  + \hat{\mu}^{(1)}, 
\end{align}
where $\hat{\mu}^{(0)}$ and $\hat{\mu}^{(1)}$ denote the leading
and rotational $1/N_c$ contributions, and the linear $m_{\mathrm{s}}$
corrections respectively 
\begin{align}
\hat{\mu}^{(0)} & =  
\;\;w_{1}D_{\mathcal{Q}3}^{(8)}
\;+\;w_{2}d_{pq3}D_{\mathcal{Q}p}^{(8)}\cdot\hat{J}_{q}
\;+\;\frac{w_{3}}{\sqrt{3}}D_{\mathcal{Q}8}^{(8)}\hat{J}_{3},\cr
\hat{\mu}^{(1)} & =  
\;\;\frac{w_{4}}{\sqrt{3}}d_{pq3}D_{\mathcal{Q}p}^{(8)}D_{8q}^{(8)}
+w_{5}\left(D_{\mathcal{Q}3}^{(8)}D_{88}^{(8)}+D_{\mathcal{Q}8}^{(8)}D_{83}^{(8)}\right)
\;+\;w_{6}\left(D_{\mathcal{Q}3}^{(8)}D_{88}^{(8)}-D_{\mathcal{Q}8}^{(8)}D_{83}^{(8)}\right).
\label{eq:magop}
\end{align}
 $d_{pq3}$ is the SU(3) symmetric tensor of which the indices run over 
$p=4,\cdots,\,7$. $\hat{J_3}$ and $\hat{J}_{p}$ denote the third
and the $p$th components of the spin operator acting on the soliton
with the light-quark pair. $D_{\mathcal{Q}3}^{(8)}$ arises from the
rotation of the electromagnetic current  
\begin{align}
D_{\mathcal{Q}3}^{(8)} = \frac12 \left( D_{33}^{(8)} + \frac1{\sqrt{3}}
  D_{83}^{(8)}\right).
\end{align}
The coefficients $w_i$ in Eq.~\eqref{eq:magop} are independent of
baryons involved, which encode the interaction of light quarks with
the electromagnetic current. Each term has a physical meaning: $w_1$
represents the leading-order contribution, a part of the rotational
$1/N_c$ corrections, and linear $m_{\mathrm{s}}$ corrections, whereas
$w_2$ and $w_3$ describe the rest of the rotational $1/N_c$
corrections. $w_1$ includes the $m_s$-dependent term, which is not
explicitly involved in the breaking of flavor SU(3) symmetry. So, we
need to treat $w_1$ as if it had contained the SU(3) symmetric part.
 On the other hand, $w_4$, $w_5$, and $w_6$ are the
SU(3) symmetry breaking terms. There are yet another $m_{\mathrm{s}}$
corrections, which arise from the collective wave functions. Though 
$w_i$ can be determined within a specific chiral solitonic model such
as the $\chi$QSM~\cite{Kim:1995mr, Kim:1995ha}, we will use the values
of $w_i$, which have been already fixed from the experimental data on
the magnetic moments of the baryon octet. 

The baryon wave function given in Eq.~\eqref{eq:waveftn} is not enough
to compute the magnetic moments, because the collective wave functions
should be revised when the perturbation coming from the strange
current quark mass is considered. In this case, the baryon is no more
in a pure state but is mixed with higher representations. 
In Ref.~\cite{Kim:2018nqf}, the collective baryon wave functions for
the heavy baryons have been already derived. Those for the baryon
anti-triplet ($J_{L}=0$) and the sextet ($J_{L}=1$) are expressed respectively
as~\cite{Kim:2018nqf}    
\begin{align}
&|B_{\overline{\bm3}_{0}}\rangle = |\overline{\bm3}_{0},B\rangle + 
p^{B}_{\overline{15}}|\overline{\bm{15}}_{0},B\rangle, \cr
&|B_{\bm6_{1}}\rangle = |{\bm6}_{1},B\rangle +
  q^{B}_{\overline{15}}|{\overline{\bm{15}}}_{1},B 
\rangle + q^{B}_{\overline{24}}|{
{\overline{\bm{24}}}_{1}},B\rangle,
\label{eq:mixedWF1}
\end{align}
with the mixing coefficients
\begin{eqnarray}
p_{\overline{15}}^{B}
\;\;=\;\;
p_{\overline{15}}\left[\begin{array}{c}
-\sqrt{15}/10\\
-3\sqrt{5}/20
\end{array}\right], 
& 
q_{\overline{15}}^{B}
\;\;=\;\;
q_{\overline{15}}\left[\begin{array}{c}
\sqrt{5}/5\\
\sqrt{30}/20\\
0
\end{array}\right], 
& 
q_{\overline{24}}^{B}
\;\;=\;\;
q_{\overline{24}}\left[\begin{array}{c}
-\sqrt{10}/10\\
-\sqrt{15}/10\\
-\sqrt{15}/10
\end{array}\right],
\label{eq:pqmix}
\end{eqnarray}
respectively, in the basis $\left[\Lambda_{Q},\;\Xi_{Q}\right]$ for
the anti-triplet and $\left[\Sigma_{Q}\left(\Sigma_{Q}^{\ast}\right),\;
  \Xi_{Q}^{\prime}\left(\Xi_{Q}^{\ast}\right),\;\Omega_{Q}
  \left(\Omega_{Q}^{\ast}\right)\right]$ for the sextets. The
parameters $p_{\overline{15}}$, $q_{\overline{15}}$, and
$q_{\overline{24}}$ are written by 
\begin{eqnarray}
p_{\overline{15}}
\;\;=\;\;
\frac{3}{4\sqrt{3}}\overline{\alpha}{I}_{2}, 
& 
q_{\overline{15}}
\;\;=\;\;
{\displaystyle -\frac{1}{\sqrt{2}}
\left(\overline{\alpha}+\frac{2}{3}\gamma\right)
I_{2}}, 
& 
q_{\overline{24}}\;\;=\;\;
\frac{4}{5\sqrt{10}}
\left(\overline{\alpha}-\frac{1}{3}\gamma\right)
I_{2}.
\label{eq:pqmix2}
\end{eqnarray}
 Combining
Eq.~\eqref{eq:mixedWF1} with the heavy-quark spinor as in
Eq.~\eqref{eq:waveftn}, one can construct the collective wave
functions for the heavy baryon states~\cite{Yang:2018uoj}. 

Computing the baryon matrix elements of $\hat{\mu}$ in
Eq.~\eqref{eq:MagMomOp}, we get the magnetic moments of the 
heavy baryons  
\begin{equation}
\mu_{B}=\mu_{B}^{(0)}+\mu_{B}^{(\mathrm{op})}+\mu_{B}^{(\mathrm{wf})}
\label{eq:mu_B}
\end{equation}
where $\mu_{B}^{(0)}$ is the part of the magnetic moment in
the chiral limit and $\mu_{B}^{(\mathrm{op})}$ comes from
$\hat{\mu}^{(1)}$ in Eq.~\eqref{eq:MagMomOp}, which include $w_4$,
$w_5$, and $w_6$. $\mu_{B}^{(\mathrm{wf})}$ is derived from the
interference between the $\mathcal{O}(m_{\mathrm{s}})$ and
$\mathcal{O}(1)$ parts of the collective wave functions in
Eq.~\eqref{eq:mixedWF1}.   

Since the soliton with the light-quark pair for the baryon
anti-triplet has spin $J_L=0$, , the magnetic moments of the baryon
anti-triplet vanish. In this case $1/m_Q$ contributions are the
leading ones. However, we will not include them, since we need to go
beyond the mean-field approximation to consider the $1/m_Q$
contributions within the present framework. 

Since $w_1$ contains both the leading-order contributions and the 
$1/N_c$ rotational corrections, we have to decompose them. Following
the argument of Ref.~\cite{Kim:2017khv}, we can separately consider
each contribution. The coefficients $w_1$, $w_2$, and $w_3$ are
expressed in terms of the model dynamical parameters  
\begin{align}
  \label{eq:w123}
w_{1}  =  
M_{0}\;-\;\frac{M_{1}^{\left(-\right)}}{I_{1}^{\left(+\right)}},\;\;\;
w_{2}  = -2\frac{M_{2}^{\left(-\right)}}{I_{2}^{\left(+\right)}},\;\;\;
w_{3}  = -2\frac{M_{1}^{\left(+\right)}}{I_{1}^{\left(+\right)}},
\end{align}
where the explicit forms of $M_0$, $M_1^{(\pm)}$, $M_2^{(-)}$ are
given in Refs.~\cite{Kim:1995mr, Praszalowicz:1998j}. $I_1^{(+)}$ and 
$I_2^{(+)}$ are the moments of inertia with the notation of
Ref.~\cite{Praszalowicz:1998j} taken. In the limit of the small soliton
size, the parameters in Eq.~\eqref{eq:w123} can be simplified as  
\begin{align}
M_{0}\;\rightarrow\;-2N_{c}K,
\;\;\;
\frac{M_{1}^{\left(-\right)}}{I_{1}^{\left(+\right)}}
\;\rightarrow\;\frac{4}{3}K, \;\;\;
   \frac{M_{1}^{\left(+\right)}}{I_{1}^{\left(+\right)}}
  \;\rightarrow\;-\frac{2}{3}K,\;\;\;
\frac{M_{2}^{\left(-\right)}}{I_{2}^{\left(+\right)}}
\;\rightarrow\;-\frac{4}{3}K.
\label{eq:sss}  
\end{align} 
These results yield the expressions of the magnetic moments in the
nonrelativistic (NR) quark model. For example, the ratio of the proton
and magnetic moments can be correctly obtained as
$\mu_p/\mu_n=-3/2$. In the NR limit, we also derive the relation
$M_{1}^{\left(-\right)}\;=\;-2M_{1}^{\left(+\right)}$. Furthermore, we
have to assume that this relation can be also applied to the 
case of the realistic soliton size. Then, we can write the 
leading-order contribution $M_0$ in terms of $w_1$ and $w_3$
\begin{align}
  \label{eq:4}
M_0= w_1 + w_3.  
\end{align}
Since a heavy baryon constitutes $N_c-1$ valence quarks, the original
$M_0$ is modified by introducing $(N_c-1)/N_c$. As mentioned
previously, only the valence part of $M_0$ should be changed by this
scaling factor. Since, however, we have determined the values of
$w_i$ using the experimental data, we can not fix separately the
valence and sea parts. Thus, we introduce an additional scaling factor
$\sigma$ to express a new coefficient $\tilde{w}_1$ 
\begin{align}
\label{eq:w1tilde}
\tilde{w}_1 = \left[\frac{N_c-1}{N_c} (w_1+w_3) - w_3\right] \sigma. 
\end{align}
$\sigma$ compensates also possible deviations from the NR relation
$M_{1}^{\left(-\right)}\;=\;-2M_{1}^{\left(+\right)}$ assumed to be 
valid in the realistic soliton case. The value of $\sigma$ is taken to
be $\sigma\sim0.85$. 

Considering the scaling parameters, we are able to determine the
following values for $w_i$
\begin{align}
\tilde{w_{1}} 
& =  
-10.08\pm0.24,
\cr
w_{2} & =  4.15\pm0.93,
\cr
w_{3} & =  8.54\pm0.86,
\cr
\overline{w}_{4} & =  -2.53\pm0.14,
\cr
\overline{w}_{5} & =  -3.29\pm0.57,
\cr
\overline{w}_{6} & =  -1.34\pm0.56.
\label{eq:numW}
\end{align}

Before we carry on the calculation of the magnetic moments, we examine
the general relations between them. First, we find the generalized
Coleman and Glashow relations~\cite{Coleman:1961jn}, which arise from 
the isospin invariance
\begin{align}
\mu(\Sigma_{c}^{++})\;-\;\mu(\Sigma_{c}^{+})
& =  
\mu(\Sigma_{c}^{+})\;-\;\mu(\Sigma_{c}^{0}),
\cr
\mu(\Sigma_{c}^{0})\;-\;\mu(\Xi_{c}^{\prime0}) 
& =  
\mu(\Xi_{c}^{\prime0})\;-\;\mu(\Omega_{c}^{0}),
\cr
2 [\mu(\Sigma_{c}^{+})\,-\,\mu(\Xi_{c}^{\prime0})]
& =  
\mu(\Sigma_{c}^{++})\,-\,\mu(\Omega_{c}^{0}).
\label{eq:coleman}  
\end{align}
Similar relations were also found in Ref.~\cite{Banuls:1999mu}.
However, there is one very important difference. While the 
Coleman-Glashow relations are known to be valid in the chiral 
limit, the relations in Eq.~\eqref{eq:coleman} are justified 
even when the effects of SU(3) flavor symmetry breaking are
considered. We also find the relation according to
the $U$-spin symmetry
\begin{align}
\mu(\Sigma_{c}^{0})\;=\;
\mu(\Xi_{c}^{\prime0})\;=\;
\mu(\Omega_{c}^{0})\;=\;
-2\mu(\Sigma_{c}^{+})\;=\;
-2\mu(\Xi_{c}^{\prime+})\;=\;
-\frac{1}{2}\mu (\Omega_{c}^{0}),
\label{eq:Usym}
\end{align}
which are only valid in the SU(3) symmetric case. 
We derive also the sum rule given as    
\begin{align}
\sum_{B_c\in\mathrm{sextet}}\mu(B_c)\;=\;0
\label{eq:sum}
\end{align}
in the SU(3) symmetric case. 

\begin{table}[htp]
\caption{Numerical results of the magnetic moments for the charmed
  baryon sextet with $J=1/2$ in units of the nuclear magneton $\mu_N$.} 
\renewcommand{\arraystretch}{1.3}% Wider\centering{}
\begin{tabular}{ccc}
\hline \hline
$\mu\left[6_{1}^{1/2},\;B_{c}\right]$ 
& $\mu^{(0)}$ 
& $\mu^{(\text{total})}$ 
\tabularnewline \hline
$\Sigma_{c}^{\text{++}}$ 
& $2.00\pm0.09$
& $2.15\pm0.1$ 
\tabularnewline
$\Sigma_{c}^{\text{+}}$ 
& $0.50\pm0.02$ 
& $0.46\pm0.03$ 
\tabularnewline
$\Sigma_{c}^{0}$ 
& -$1.00\pm0.05$ 
& -$1.24\pm0.05$ 
\tabularnewline
\hline 
$\Xi_{c}^{\prime+}$ 
& $0.50\pm0.02$ 
& $0.60\pm0.02$ 
\tabularnewline
$\Xi_{c}^{\prime0}$ 
& -$1.00\pm0.05$ 
& -$1.05\pm0.04$ 
\tabularnewline
\hline 
$\Omega_{c}^{0}$ 
& -$1.00\pm0.05$ 
& -$0.85\pm0.05$ 
\tabularnewline
\hline \hline
\end{tabular}
\label{tab:3}
\end{table}
\begin{table}[htp]
\renewcommand{\arraystretch}{1.3}% Wider\centering{}
\caption{Numerical results of magnetic moments for charmed baryon
  sextet with $J=3/2$ in units of the nuclear magneton $\mu_N$.} 
\begin{tabular}{ccc}
\hline \hline 
$\mu\left[6_{1}^{3/2},\;B_{c}\right]$ 
& $\mu^{(0)}$ 
& $\mu^{(\text{total})}$ 
\tabularnewline
\hline 
$\Sigma_{c}^{\ast\text{++}}$ 
& $3.00\pm0.14$ 
& $3.22\pm0.15$ 
\tabularnewline
$\Sigma_{c}^{\ast\text{+}}$ 
& $0.75\pm0.04$ 
& $0.68\pm0.04$ 
\tabularnewline
$\Sigma_{c}^{\ast0}$ 
& $-1.50\pm0.07$ 
& $-1.86\pm0.07$ 
\tabularnewline
\hline 
$\Xi_{c}^{\ast+}$ 
& $0.75\pm0.04$ 
& $0.90\pm0.04$ 
\tabularnewline
$\Xi_{c}^{\ast0}$ 
& $-1.50\pm0.07$ 
& $-1.57\pm0.06$ 
\tabularnewline
\hline 
$\Omega_{c}^{\ast0}$ 
& -$1.50\pm0.07$ 
& -$1.28\pm0.08$ 
\tabularnewline
\hline \hline
\end{tabular}
\label{tab:4}
\end{table}
In Tables~\ref{tab:3} and \ref{tab:4}, we list the numerical results
of the charmed baryon sextet with spin 1/2 and 3/2, respectively. We
obtain exactly the same results for the bottom baryons because of the
heavy-quark symmetry in the $m_Q\to\infty$ limit. In
Ref.~\cite{Yang:2018uoj}, a detailed discussion can be found, the
present results being compared with those from many other models. 
\section{Excited $\Omega_c$ baryons}
The present mean-field approach was applied to the classification of
the excited $\Omega_c^0$'s that were recently reported by the LHCb
Collaboration~\cite{Aaij:2017nav}. The masses and decay widths of the
$\Omega_c^0$'s, which were reported by the LHCb Collaboration, are
listed in Table~\ref{tab:5}. The Belle Collaboration has confirmed the
four of them~\cite{Yelton:2017qxg} (see Table~\ref{tab:6}). The Belle
data unambiguously confirmed the existence of the $\Omega_c(3066)$ and
$\Omega_c(3090)$, and $\Omega_c(3000)$ and $\Omega_c(3050)$ are also
confirmed with reasonable significance. On the other hand the narrow
resonance $\Omega_c(3119)$ was not seen in the Belle experiment but
the nonobservation of $\Omega_c(3119)$ is not in disagreement because
it is due to the small yield. 
\begin{table}[htp]
\renewcommand{\arraystretch}{1.3}% Wider\centering{}
\caption{Experimental data on the five $\Omega_c^0$ baryons reported by the
  LHCb Collaboration~\cite{Aaij:2017nav}.} 
\begin{tabular}{ccc}
\hline \hline 
Resonance
& Mass (MeV)
& Decay width  (MeV)
\tabularnewline
\hline 
 $\Omega_c(3000)^0$
& $3000.4\pm 0.2\pm0.1_{-0.5}^{+0.3}$ 
& $4.5\pm 0.6\pm 0.3$
\tabularnewline
 $\Omega_c(3050)^0$
& $3050.2\pm0.1\pm0.1_{-0.5}^{+0.3}$
& $0.8\pm 0.2\pm 0.1$
\tabularnewline
 $\Omega_c(3066)^0$
& $3065.6\pm 0.1\pm 0.3_{-0.5}^{+0.3}$ 
& $3.5\pm0.4\pm0.2$
\tabularnewline
 $\Omega_c(3090)^0$
& $3090.2\pm 0.3\pm 0.5_{-0.5}^{+0.3}$
& $8.7\pm 1.0 \pm 0.8$
\tabularnewline
 $\Omega_c(3119)^0$
& $3119.1\pm 0.3 \pm 0.9_{-0.5}^{+0.3}$
& $1.1 \pm 0.8 \pm 0.4$
\tabularnewline
$\Omega_{c}(3188)$ 
& $3188\pm 5 \pm 13$
& $60\pm 15 \pm 11$
\tabularnewline
\hline \hline
\end{tabular}
\label{tab:5}
\end{table}

\begin{table}[htp]
\renewcommand{\arraystretch}{1.3}% Wider\centering{}
\caption{Experimental data on the four $\Omega_c^0$ baryons reported by the
  Belle Collaboration~\cite{Yelton:2017qxg}.}
\begin{tabular}{cc}
\hline \hline 
Resonance
& Mass (MeV)
\tabularnewline
\hline 
 $\Omega_c(3000)^0$
& $3000.7\pm 1.0\pm 0.2$
\tabularnewline
 $\Omega_c(3050)^0$
& $3050.2\pm0.4\pm0.2$
\tabularnewline
 $\Omega_c(3066)^0$
& $3064.9\pm 0.6\pm 0.2$
\tabularnewline
 $\Omega_c(3090)^0$
& $3089.3\pm 1.2\pm 0.2$
\tabularnewline
 $\Omega_c(3119)^0$
& --
\tabularnewline
$\Omega_{c}(3188)$ 
& $3199\pm 9 \pm 4$
\tabularnewline
\hline \hline
\end{tabular}
\label{tab:6}
\end{table} 

When one examines the excited heavy baryons in the present work, we
need to consider states with the grand spin $K=1$. Since we have the
quantization rule $\bm{K}=\bm{J}_L+\bm{T}$, the possible values of the
spin are determined by 
\begin{align}
J_L = |K-T|,\cdots, K+T.  
\end{align}
Thus, In the case of $T=0$ which corresponds to the anti-triplet with
$Y'=2/3$, we must have $J_L=1$ because of $K=1$. Combining it with the
heavy-quark spin 1/2, we have \textit{two} excited baryon
anti-triplet. Similarly, $T=1$ corresponds to the sextet. In this
case $J_L$ can have the values of 0, 1, and 2. Being coupled with the
heavy-quark spin $1/2$, we get \textit{five} excited baryon sextets:
$(1/2)$, $(1/2,\,3/2)$, and $(3/2,\,5/2)$, corresponding to $J_L=0$, and
$J_L=1$, and $J_L=2$. In each sextet representation, we have a
isosinglet $\Omega_c^0$. Thus, is is natural to think that the newly
found five $\Omega_c^0$'s are those in the excited baryon sextets.   
Note that the representations for each value of $J$ are degenerate in
the limit of $m_Q\to\infty$. So, we need to introduce an additional
hyperfine spin-spin interaction as done for the ground-state baryon
sextet
\begin{align}
H_{LQ}  = \frac23 \frac{\varkappa'}{m_Q} \bm{J}_L \cdot \bm{J}_Q,
\end{align}
which is very similar to Eq.~\eqref{eq:ssinter}. $\varkappa'$ can be
fixed by using the experimental data on the masses of the excited
baryon anti-triplet.

Following Refs.~\cite{Diakonov:2012zz, Diakonov:2013qta}, we revise the
eigenvalues of the symmetric Hamiltonian for the excited baryons
($K\neq 0$) as 
follows 
\begin{align}
M_{\mathcal{R}}^{(K)\prime} &=  M_{\mathrm{cl}}^{(K)\prime} + \frac1{2I_2} \left[
  C_2(\mathcal{R}) - T (T+1) - \frac34 Y^{\prime 2} \right]  \cr
& + \frac1{2I_1} \left[(1-a_K) T(T+1) + a_K J_L(J_L+1) - a_K(1-a_K) K(K+1) 
  \right], 
\label{eq:symmass}
\end{align}
where $C_2(\mathcal{R})$ is the eigenvalue of the SU(3) Casimir
operator, which was already defined in Eq.~\eqref{eq:Casimir}.  The
parameter $a_K$ is related to one-quark excitation. 
The collective wave functions for the soliton are derived as  
\begin{align}
\Phi_{B,J_L, J_{L3},(T,K)}^{\mathcal{R}} = \sqrt{\frac{2J_L+1}{2K+1}} \sum_{T_3
  J_{L3}' K_3'} C_{TT_3J_LJ_{L3}'}^{KK_3}  (-1)^{(T+T_3)}
  \Psi_{(\mathcal{R^*};-Y'TT_3)}^{(\mathcal{R};B)} D_{J_{L3}'J_{L3}}^{(J_L)*}(S)
  \chi_{K_3'}, 
\end{align}
where index $(\mathcal{R};YTT_3)$ denotes the SU(3) quantum numbers of
a corresponding baryon in representation $\mathcal{R}$, and
$(\mathcal{R}^*;-Y'TT_3)$ is attached to a fixed value of $Y'$ and
is formally given in a conjugate representation to $\mathcal{R}$. The
function $D^{(J_L)}$ represents the SU(2) Wigner $D$ function and
$\chi_{K_3}$ is the spinor corresponding to $K$ and $K_3$. The 
wave function for the excited baryons can be constructed by coupling
$\Phi_{B,J_L, J_{L3},(T,K)}^{\mathcal{R}}$ with the heavy-quark
spinor.  

The SU(3) symmetry-breaking Hamiltonian in Eq.~\eqref{sb} also needs to
be extended to describe the mass splittings of the excited heavy
baryons 
\begin{align}
H_{\mathrm{sb}}^{(K)} = \overline{\alpha} D_{88}^{(8)} + \beta \hat{Y}
  + \frac{\gamma}{\sqrt{3}} \sum_{i=1}^3 D_{8i}^{(8)} \hat{T}_i +
  \frac{\delta}{\sqrt{3}} \sum_{i=1}^3 D_{8i}^{(8)} \hat{K}_i. 
\label{eq:excitedsu3br}
\end{align}
The additional parameter $\delta$ can be determined by using the mass
spectrum of excited baryons. 

\begin{figure}[htp]
\centering
\includegraphics[scale=0.2]{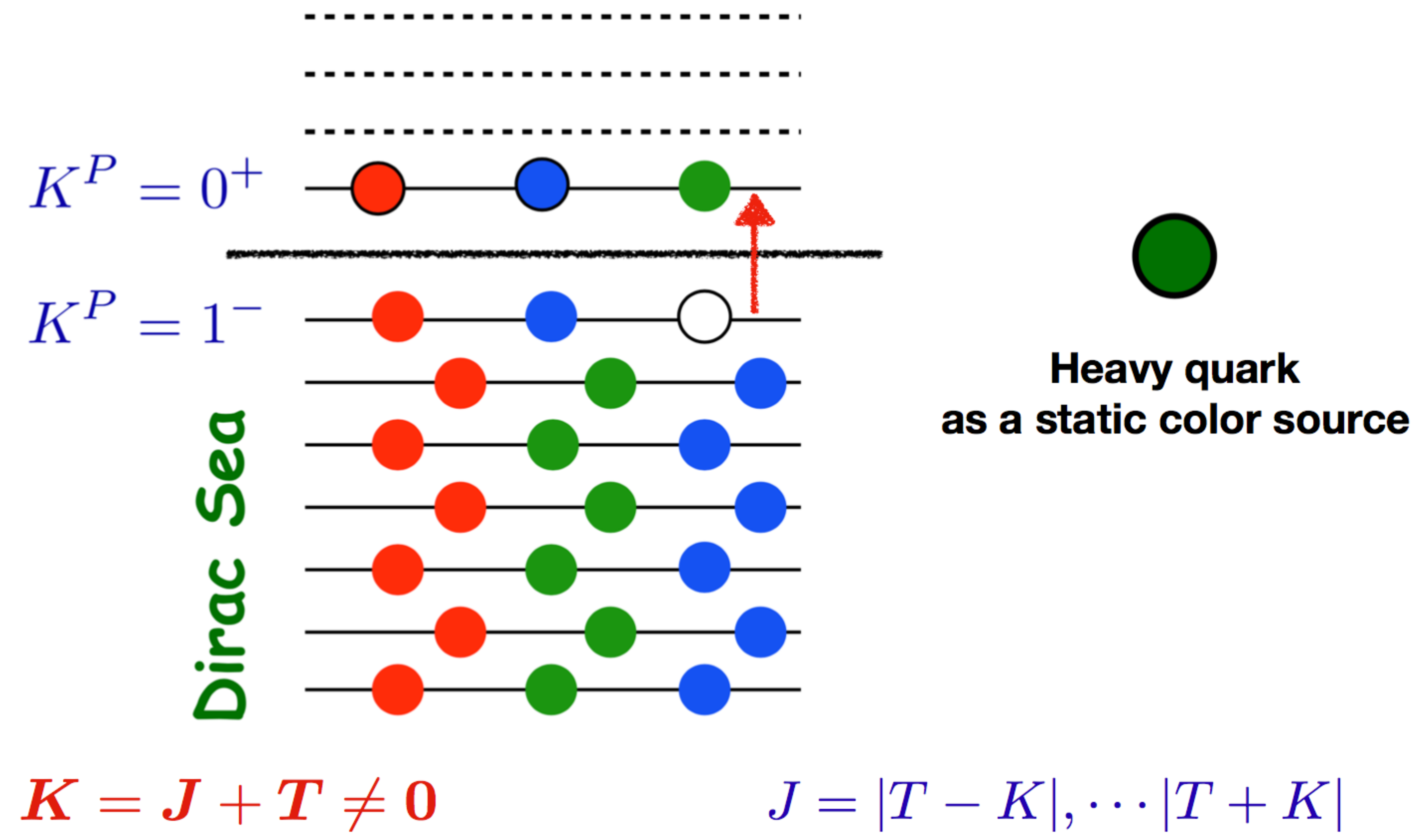}\qquad
\caption{Schematic picture of the first excited heavy baryons. A
  possible excitation of a quark from the Dirac sea to the valence
  level might have $K^P=1^-$.} 
\label{fig:5}
\end{figure}
As shown in Fig.~\ref{fig:5}, the transition from a $K^P=1^-$
Dirac-sea level to an unoccupied $K^P=0^+$ state may correspond to the
first excited heavy baryons~\cite{Diakonov:2010tf}. Note that such a transition
is only allowed in the heavy-baryon sector, not in the light-baryon
sector. As discussed already, there are two baryon anti-triplets and
five baryon sextets. From Eq.~\eqref{eq:symmass}, we can derive the
following expressions
\begin{align}
M_{\overline{\bm{3}}}^{\prime} &= M_{\mathrm{cl}}^{\prime} +
  \frac1{2I_2} + \frac1{I_1} (a_1^2),\cr
M_{\bm{6}}^{\prime} &= M_{\overline{\bm{3}}}^{\prime} +
                           \frac{1-a_1}{I_1} + \frac{a_1}{I_1}\times
                          \left\{
                           \begin{array}{ll} 
-1 & \mbox{ for $J_{L}=0$} \\
0 & \mbox{ for $J_{L}=1$} \\
2 & \mbox{ for $J_{L}=2$}
                           \end{array} \right. .
  \label{eq:excitedM3-6}
\end{align}
Considering the SU(3) symmetry breaking from
Eq.~\eqref{eq:excitedsu3br}, we find the splitting parameters for the
$\overline{\bm{3}}$ and $\bm{6}$ 
\begin{align}
\delta_{\overline{\bm{3}}}' &= \frac38 \overline{\alpha} + \beta =
  \delta_{\overline{\bm{3}}} = -180\,\mathrm{MeV},\cr
 \delta_{\bm{6}J_{L}}' &= \delta_{\bm{6}} -\frac{3}{20}\delta \times
                     \left\{
 \begin{array}{ll} 
-1 & \mbox{ for $J_{L}=0$} \\
0 & \mbox{ for $J_{L}=1$} \\
2 & \mbox{ for $J_{L}=2$}
                           \end{array} \right.,
\label{eq:excitedeltas}
\end{align}
where we see that $\delta_{\overline{\bm{3}}}'$ is just the same as
$\delta_{\overline{\bm{3}}}$ given in
Eq.~\eqref{eq:deltas}. $\delta_{\bm{6}}$ is given as $-120$
MeV. Though we do not know the numerical value of the new parameter
$\delta$, we still can analyze the mass splittings of the newly found
$\Omega_c$'s, using the splittings between the states with different
values of $J_{L}$. 

\begin{figure}[htp]
\centering
\includegraphics[scale=0.2]{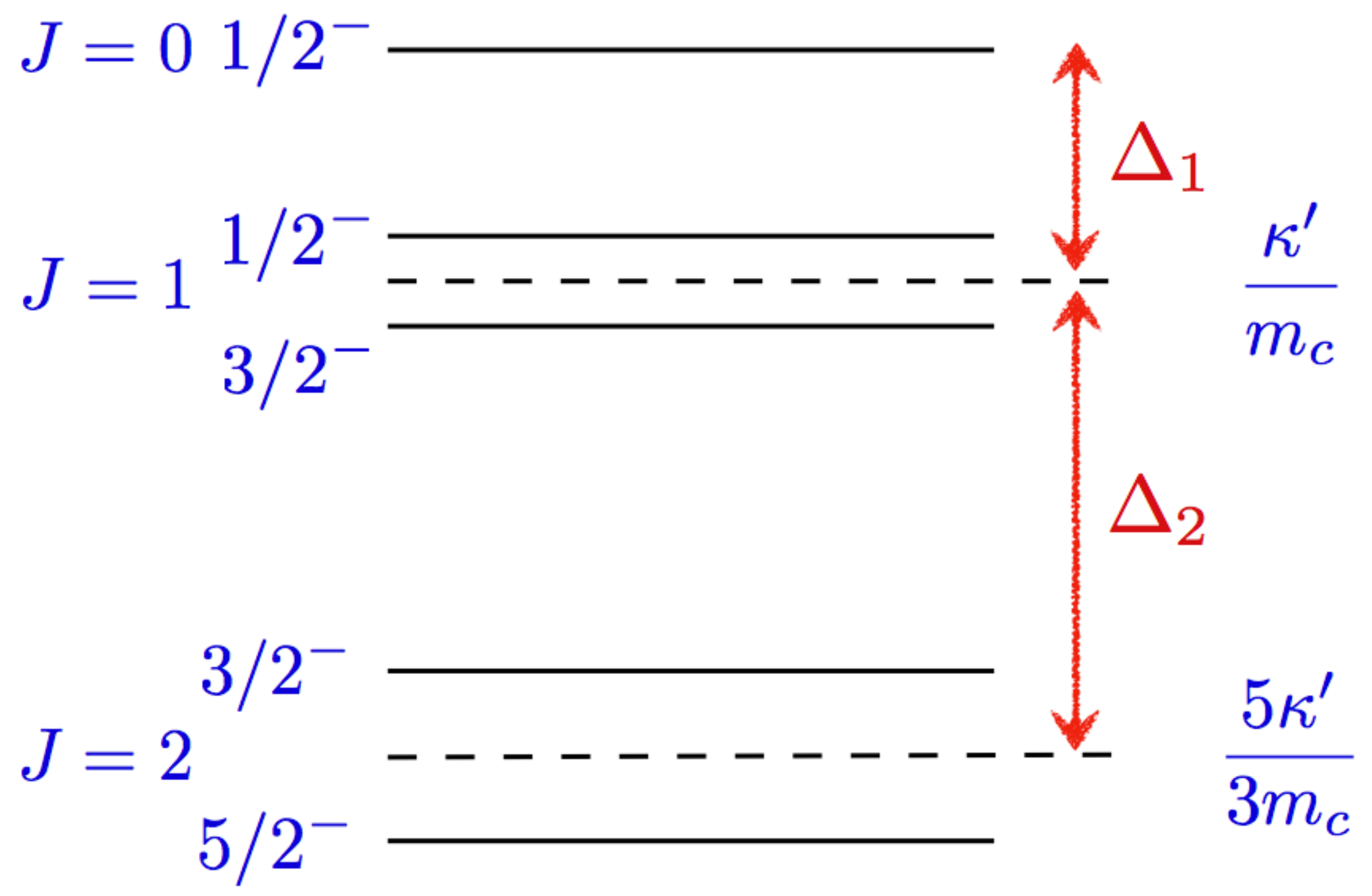}\qquad
\caption{Mass splitting of the five excited sextets.}
\label{fig:6}
\end{figure}

We now turn to the hyperfine splittings. The two anti-triplets of spin
1/2 and 3/2 and the two sextets of spin 1/2 and 3/2 are split by 
\begin{align}
\Delta_{\overline{\bm{3}}}^{\mathrm{hf}} =
  \Delta_{\bm{6}J_{L}=1}^{\mathrm{hf}} = \frac{\varkappa'}{m_c},    
\end{align}
whereas another two sextets of spin 3/2 and 5/2 are split by
\begin{align}
  \Delta_{\bm{6}J_{L}=2}^{\mathrm{hf}} = \frac53 \frac{\varkappa'}{m_c}.    
\label{eq:hf6}
\end{align}
One sextet of spin 1/2 from the $J_{L}=0$ case has no hyperfine
splitting. The results are depicted in Fig.~\ref{fig:6}.
Note that the $\Delta_1$ represent the splittings
between the $J_{L}=0$ state and the degenerate $J_{L}=1$ state, whereas
$\Delta_2$ denote those between degenerate $J_{L}=1$ and $J_{L}=2$ states 
\begin{align}
\Delta_1 = \frac{a_1}{I_1} + \frac{3}{20} \delta,\;\;\; \Delta_2 =
  2\Delta_1.  
\label{eq:Jsplit}
\end{align}
We will soon see that the relation $\Delta_1=2\Delta_2$ will play an
critical role 
in identifying the excited $\Omega_c$'s within the $\chi$QSM. 

If one identifies $\Lambda_c(2592)$ and $\Xi_c(2790)$ as the members
of the excited baryon anti-triplet of spin $(1/2)^-$ with negative
parity, and $\Lambda_c(2592)$ and $\Xi_c(2790)$ as those of the
excited baryon anti-triplet of spin $(3/2)^-$, then we find
$\delta_{\overline{\bm{3}}}=-198$ and $-190$ MeV, which are more or
less in agreement with the value given in
Eq.~\eqref{eq:excitedeltas}. The $\varkappa'/m_c$ can be also
determined as 
\begin{align}
\frac{\varkappa'}{m_c} = \frac13 (M_{\Lambda_c(2628)} + 2
  M_{\Xi_c(2818)}) - \frac13 (M_{\Lambda_c(2592)} + 2 M_{\Xi_c(2790)})
  = 30\,\mathrm{MeV}, 
\label{eq:exhfvalue}
\end{align}
and $M_{\overline{\bm{3}}}$ is also fixed by 
\begin{align}
M_{\overline{\bm{3}}} = \frac29 (M_{\Lambda_c(2628)} + 2
  M_{\Xi_c(2818)}) + \frac19 (M_{\Lambda_c(2592)} + 2 M_{\Xi_c(2790)})
  = 2744\,\mathrm{MeV}.  
\end{align}

We now assert that as a minimal scenario the newly found $\Omega_c$ baryons
by the LHCb Collaboration belong to the five excited sextets. Then
$\Omega_c(3000)$ can be identified as the state with $(J_{L}=0,\,1/2^-)$,
which corresponds to the lightest state in Fig.~\ref{fig:6}. All other
four states can be consequently identified as depicted in
Fig.~\ref{fig:6}. Including the hyperfine interactions, we get the
results as summarized in Table~\ref{tab:7}.
\renewcommand{\arraystretch}{1.3}
\begin{table}[thp]
\caption{Scenario 1: All five LHCb
  $\Omega_{c}$ states are assigned to the excited baryon sextets.}% 
\label{tab:7}%
\begin{center}%
\begin{tabular}
[c]{ccccc}\hline\hline
$J_{L}$ & $S^{P}$ & $M$~[MeV] & $\varkappa^{\prime}/m_{c}$~[MeV] &
                                                            $\Delta_{J_{L}}$~[MeV]\\ \hline
0 & $\frac{1}{2}^{-}$ & 3000 & not applicable & not applicable\\ 
\multirow{2}{*}{1} & $\frac{1}{2}^{-}$ & 3050 & \multirow{2}{*}{16} &
\multirow{2}{*}{61}\\
~ & $\frac{3}{2}^{-}$ & 3066 &  & \\ %\cline{1-5}%
\multirow{2}{*}{2} & $\frac{3}{2}^{-}$ & 3090 & \multirow{2}{*}{17} &
\multirow{2}{*}{47}\\
& $\frac{5}{2}^{-}$ & 3119 &  & \\\hline \hline
\end{tabular}
\end{center}
\par
\end{table}
\renewcommand{\arraystretch}{1}
We find at least three different contradictions arising from the
assignment of these $\Omega_c$ states as the members of the
excited sextets within the $\chi$QSM.  Firstly, this assignment
requires that the hyperfine splitting should be almost as twice as 
smaller than in the $\overline{\bm{3}}$ case. Secondly,  the robust
relation $\Delta_2=2\Delta_1$ given in Eq.~\eqref{eq:Jsplit}  is badly
broken. Finally, there are two orthogonal sum rules
$\sigma_1=\sigma_2=0$ derived from the $\chi$QSM  
\begin{align}
\sigma_1&=6\; \Omega_c(J_{L}=0,1/2^-)- \Omega_c(J_{L}=1,1/2^-)-8 \;
          \Omega_c(J_{L}=1,3/2^-)+3\; \Omega_c(J_{L}=2,5/2^-) , \label{sr}\\ 
\sigma_2&=-4\; \Omega_c(J_{L}=0,1/2^-)+9\; \Omega_c(J_{L}=1,1/2^-)-3 \;
          \Omega_c(J_{L}=1,3/2^-)-5 \; \Omega_c(J_{L}=2,3/2^-)  
+3\; \Omega_c(J_{L}=2,5/2^-),  \notag
\end{align}
which are also badly broken. Thus, we come to the conclusion that the
the five $\Omega_c$ baryons is unlikely to belong to the excited
sextets. A similar conclusion was drawn by
Ref.~\cite{Karliner:2017kfm} in a different theoretical framework.  
Moreover, the computed decay widths of the excited $\Omega_c$'s do not
match with the experimental data. Therefore, the first scenario is
unrealistic in the present mean-field approach. 

Since the first scenario is not suitable for identifying the five
excited $\Omega_c$ baryons, we have to come up with another
scenario. Observing that two of them have rather narrower decay widths
than other three $\Omega_c$'s, we assert that these narrow
$\Omega_c(3050)$ and $\Omega_c(3119)$ belong to the possible exotic
anti-decapentaplet ($\overline{\bm{15}}$) which is yet another
lowest-lying allowed representation, whereas three of them belong to
the excited sextet. We find in this scenario that two other members of
the excited baryon sextet with $J_{L}=2$ have masses above the $\Xi D$
threshold at 3185 MeV. Since they have rather broad widths, they are
not clearly seen in the LHCb data and may fall into the bump
structures appearing in the LHCb data.  

\renewcommand{\arraystretch}{1.3}
\begin{table}[thp]
\caption{Scenario 2. Only three LHCb states are
assigned to {the} sextets. }%
\label{tab:8}%
\begin{center}%
\begin{tabular}
[c]{ccccc}\hline \hline
$J_{L}$& $S^{P}$ & $M$~[MeV] & $\varkappa^{\prime}/m_{c}$~[MeV] & $\Delta_{J_{L}}%
$~[MeV]\\\hline
0 & $\frac{1}{2}^{-}$ & 3000 & not applicable & not applicable\\
\multirow{2}{*}{1} & $\frac{1}{2}^{-}$ & 3066 & \multirow{2}{*}{24} &
\multirow{2}{*}{82}\\
~ & $\frac{3}{2}^{-}$ & 3090 &  & \\
\multirow{2}{*}{2} & $\frac{3}{2}^{-}$ & \emph{3222} & input & input\\
& $\frac{5}{2}^{-}$ & \emph{3262} & 24 & 164\\\hline \hline
\end{tabular}
\end{center}
\par
\end{table}
\renewcommand{\arraystretch}{1}
The results of the second scenario are summarized in Table~\ref{tab:8}
except for the $\Omega_c(3050)$ and $\Omega_c(3119)$ which will be
discussed separately.  The italic numbers correspond to the bump
structures from which $\Omega_c(3222)$ used as input. Scenario 2
provides a much more plausible prediction than scenario 1
does. Interestingly, the value of $\varkappa'/m_c\approx 24$ MeV is
closer to that determined from the excited baryon anti-triplets, given
in Eq.~\eqref{eq:exhfvalue}. Moreover, the relation
$\Delta_1=2\Delta_2$ is nicely satisfied in this scenario.  

\begin{figure}[htp]
\centering
\includegraphics[scale=0.3]{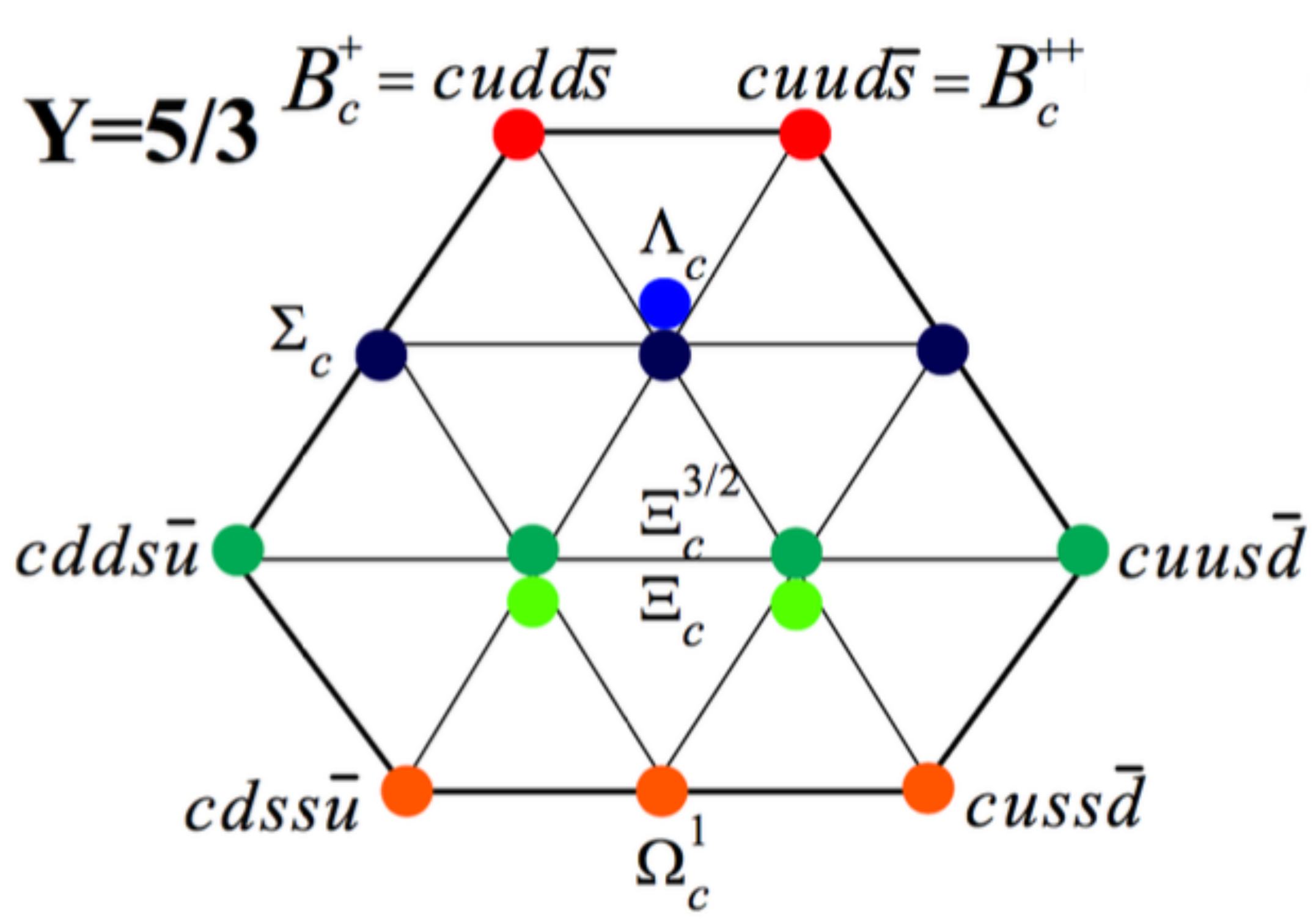}\qquad
\caption{Representation of the anti-decapentaplet
  ($\overline{\bm{15}}$). As in the case of the baryon sextet, there
  are two baryon anti-decapentaplets with spin 1/2 and 3/2. The
  $\Omega_c$s belong to the isotriplet in the $\overline{\bm{15}}$plet.}
\label{fig:7}
\end{figure}

\begin{figure}[htp]
\centering
\includegraphics[scale=0.3]{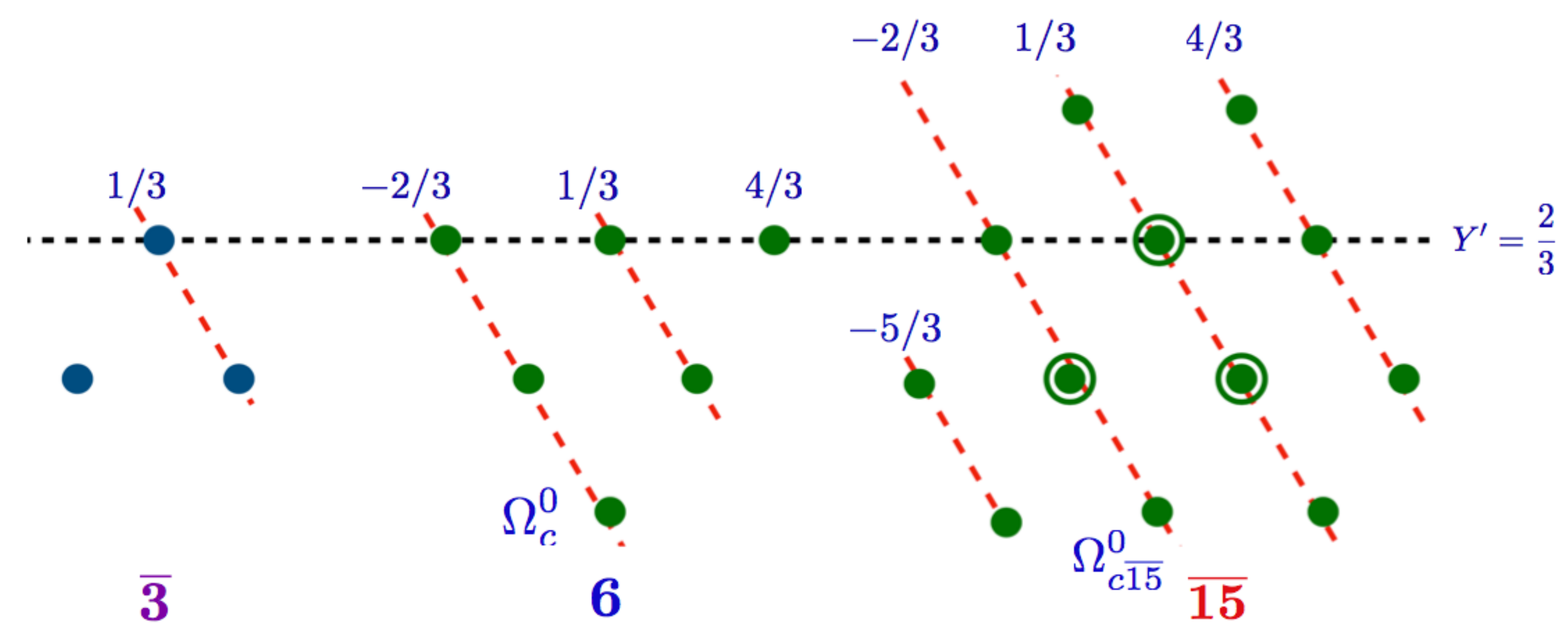}\qquad
\caption{The allowed representations for the lowest-lying heavy
  baryons. }
\label{fig:8}
\end{figure}
The anti-decapentaplet ($\overline{\bm{15}}$) was first suggested by
Diakonov~\cite{Diakonov:2010tf}. Figure~\ref{fig:7} illustrates the
representation of the $\overline{\bm{15}}$.  Since the
$\overline{\bm{15}}$ belongs to the allowed representations for the
ground-state heavy baryons, it satisfies the quantization rule
$\bm{J}_{L}+\bm{T}=\bm{0}$, so $T=J_{L}=1$ (see Fig.~\ref{fig:8}).  When the
light-quark pair with $J_{L}=1$ is coupled to the heavy-quark spin, there
are two possible $\overline{\bm{15}}$ representations that are
degenerate in the limit of $m_Q\to \infty$. It means that one needs to
consider the hyperfine interaction defined in Eq.~\eqref{eq:ssinter}.
As given in Eq.~\eqref{eq:kappavalue}, the value of $\varkappa/m_c$ is
around 68 MeV. Surprisingly, the mass difference between the
$\Omega_c(3050)$ and the $\Omega_c(3119)$ is  
\begin{align}
M_{\Omega_c(3/2^+)} (3119) -  M_{\Omega_c(1/2^+)} (3050) =
  \frac{\varkappa}{m_c} \approx   69\,\mathrm{MeV} 
\end{align}
which is almost the same as what was determined from the lowest-lying
sextet baryons. The decay widths of the excited $\Omega_c$ baryons
predicted within the present framework further support the
plausibility of scenario 2~\cite{Kim:2017khv}.  The decay widths for
the $\Omega_c(3050)$ and $\Omega_c(3119)$ are predicted to be 
\begin{align}
\Gamma_{\Omega_c(3050)(\overline{\bm{15}},1/2^+)} =
  0.48\,\mathrm{MeV}, \;\;\; 
\Gamma_{\Omega_c(3119)(\overline{\bm{15}},3/2^+)} = 1.12\,\mathrm{MeV},  
\end{align}
which are in good agreement with the LHCb data
$\Gamma_{\Omega_c(3050)}=(0.8\pm0.2\pm 0.1)$ MeV and
$\Gamma_{\Omega_c(3119)}=(1.1\pm0.8\pm 0.4)$ MeV. For detailed
discussion related to the decay widths of $\Omega_c$, we refer to
Ref.~\cite{Kim:2017khv}.   

In addition to scenarios 1 and 2, we also tried to examine
several other scenarios but find that they all turned out to be
inconsistent with the experimental data. Finally, we want to emphasize
that the $\Omega_c(3050)$ and $\Omega_c(3119)$ assigned to the members
of the $\overline{\bm{15}}$ are isotriplets. It implies that if they
indeed belong to the $\overline{\bm{15}}$, charged $\Omega_c^{\pm}$
should exist. Knowing that the excited $\Omega_c^0$'s have been
measured in the $\Xi_c^+K_c^-$ channel, we propose that the $\Xi_c^+
K^0$ and $\Xi_c^0 K^-$ channels need to be scanned in the range of the
invariant mass between 3000 MeV and 3200 MeV to find an isovector
$\Omega_c$'s. If they do not exist, this will falsify the present
predictions. 
\section{Conclusion and outlook}
In the present short review, we briefly summarized a series of recent
works on the properties of the singly heavy baryons within a pion
mean-field approach, also known as the chiral quark-soliton model.
In the limit of the infinitely heavy quark mass ($m_Q\to \infty$), 
the heavy quark inside a heavy baryon can be treated as a mere static
color source. Then a heavy baryon is portrayed as a state of $N_c-1$
valence quarks bound by the pion mean field with a heavy stripped off
from the valence level.  This mean-field approach has a certain
virtue since both the light and heavy baryons can be dealt with on an
equal footing. It means that we can bring all dynamical parameters
which have been already determined in the light-baryon sector to
describe the heavy baryons. Indeed we can simply replace the
$N_c$-counting prefactor by $N_c-1$ for the valence contributions to
the heavy baryons. Accordingly, we were able to explain the masses of
the lowest-lying heavy baryons and the magnetic moments of them
without introducing additional parameters except for the hyperfine
spin-spin interactions. We have employed the same framework to
identify the newly found excited $\Omega_c$ baryons reported by the
LHCb Collaboration. Assigning the three of them to the excited baryon
sextets and the two of them with narrower decay widths to the possible
exotic baryon anti-decapentaplet, we were able to classify the
$\Omega_c$'s successfully.  Since the $\Omega_c$ baryons in the
anti-decapentaplet are the isovector baryons, we anticipate that
charged $\Omega_c$'s might be found in other channels such as the $\Xi_c^+
K^0$ and $\Xi_c^0 K^-$. 

The present model can be further applied to future investigations on
various properties and form factors of heavy baryons. As already shown
in Ref.~\cite{Kim:2018nqf}, the electric form factor of the charged
heavy baryon indicates that a heavy baryon is an electrically compact
object. Transition form factors of heavy baryons will further reveal
their internal structure. Understanding excited heavy baryons is
another crucial issue that should be investigated. Related studies are
under way. 

\begin{acknowledgments}
I am very grateful to M. V. Polyakov, M. Prasza{\l}owicz, and
Gh.-S. Yang for fruitful collaborations and discussions over decades.  
I am thankful to J.-Y. Kim for the discussion related to the
electromagnetic form factors of heavy baryons. 
I want to express the gratitude to the editors of the Journal of the
Korean Physical Society (JKPS) for giving me an opportunity to join
the very special 50th anniversary celebration of the JKPS.   
The present work was supported by the Inha University Grant in 2017.
\end{acknowledgments}

\end{document}